\documentclass[aps,preprint,superscriptaddress,nofootinbib,11pt,]{revtex4-1}
\usepackage[pdftex]{graphicx}
\usepackage{amsmath,amssymb,color,colortbl}
\usepackage[utf8]{inputenc}
\usepackage{braket}

\begin{document}
\preprint{KANAZAWA-20-07}
\title{
	A hybrid seesaw model and hierarchical neutrino flavor structures\\  based on $A_{4}$ symmetry
}
\author{Mayumi~Aoki}
\email{mayumi@hep.s.kanazawa-u.ac.jp}
\affiliation{
Institute for Theoretical Physics, 
Kanazawa University, 
Kanazawa 920-1192, Japan
}
\author{Daiki~Kaneko}
\email{d$\_$kaneko@hep.s.kanazawa-u.ac.jp}
\affiliation{
Institute for Theoretical Physics, 
Kanazawa University, 
Kanazawa 920-1192, Japan
}
\begin{abstract}
We propose a hybrid seesaw model based on $A_{4}$ flavor symmetry, which generates a large hierarchical flavor structure. 
In our model, tree-level and one-loop seesaw mechanisms predict different flavor structures in the neutrino mass matrix and generate
a notable hierarchy among them.
We find that such a hierarchical structure gives a large effective neutrino mass that can be accessible by 
next-generation neutrinoless double beta decay
experiments.
Majorana phases can also be predictable.
The $A_{4}$ flavor symmetry in the model is spontaneously broken to the $Z_{2}$ symmetry, leading to a dark matter candidate that is assumed to be a neutral scalar field.
The favored mass region of the dark matter is obtained by numerical computations of the relic abundance and the cross section of the nucleon.
We also investigate the predictions of several hierarchical flavor structures based on $A_{4}$ symmetry
for the effective neutrino mass and the Majorana phases, and find
 characteristic features depending on the hierarchical structures.
\end{abstract}
\maketitle

\section{Introduction}
The discovery of neutrino oscillations shows that neutrinos are mixed with each other and have tiny masses.
Since neutrinos are massless particles in the Standard Model (SM), new physics beyond the SM that has some mechanism to generate 
the neutrino masses is required.
The Type-I seesaw mechanism  \cite{Minkowski:1977sc,Mohapatra:1979ia,GellMann:1980vs,Schechter:1980gr,Yanagida:1980xy} is one of the attractive ways to generate such 
tiny neutrino masses at tree level, which requires an introduction of right-handed neutrinos. 
Another attractive way to explain the tiny masses is a radiative seesaw mechanism in which neutrino masses are generated 
by loop effects (see Refs. \cite{Zee:1980ai,Zee:1985id,Babu:1988ki,Krauss:2002px,Ma:2006km,Aoki:2008av,Gustafsson:2012vj} for early works and also Ref. \cite{Cai:2017mow} for a later review).
In radiative seesaw models involving right-handed neutrinos \cite{Krauss:2002px,Ma:2006km,Aoki:2008av}  a discrete symmetry is imposed to forbid the Type-I seesaw mechanism.
This symmetry is also responsible for the stability of dark matter (DM).

The neutrino mass matrix is diagonalized by the lepton flavor mixing matrix, the so-called
Pontecorvo-Maki-Nakagawa-Sakata (PMNS) matrix, which is parameterized as 
\begin{align}
U_{\rm PMNS}^{} =&
\left(\begin{array}{ccc}
c_{12}c_{13}                        & s_{12}c_{13}                  & s_{13}e^{-i\delta} \\
-s_{12}c_{23}-c_{12}s_{23}s_{13}e^{i\delta}  & c_{12}c_{23}-s_{12}s_{23}s_{13}e^{i\delta}  & s_{23}c_{13} \\
s_{12}s_{23}-c_{12}c_{23}s_{13}e^{i\delta}   & -c_{12}s_{23}-s_{12}c_{23}s_{13}e^{i\delta} & c_{23}c_{13}  
\end{array}\right)
\left(
\begin{array}{ccc}
1 & 0 &0 \\
0 & e^{i\alpha_2/2} &0 \\
0&0&e^{i\alpha_3/2}
\end{array}
\right)\,,
 \label{UPMNS}
\end{align}
where $c_{ij}=\cos\theta_{ij}$ and $s_{ij}=\sin\theta_{ij}$ for $i, j=1,2,3$. The parameter $\delta$ is a Dirac phase, while $\alpha_2$ and $\alpha_3$ denote Majorana phases.   
The data obtained in neutrino oscillation experiments \cite{Abe:2017aap,Bak:2018ydk,Acero:2019ksn,DoubleChooz:2019qbj, Abe:2020vdv} show
that the neutrino mixing angles are 
$\theta_{12}\simeq 33^{\circ} , \theta_{23} \simeq 49^{\circ} ,\theta_{13} \simeq 8.6^{\circ} $, 
and the neutrino mass-squared differences $\Delta m_{31}^{2}\equiv m_3^2-m_1^2$ and $\Delta m_{21}^{2}\equiv m_2^2-m_1^2$ are
$|\Delta m_{31}^{2}| \simeq 2.5  \times 10^{-3}~{\rm eV}^{2}$ and $\Delta m_{21}^{2} \simeq 7.4 \times10^{-5}~{\rm eV}^{2}$, respectively.
The recent measurements of the Dirac phase show $\delta  = 107^{\circ} \to 403^{\circ}$
for the normal mass ordering (NO)
and $\delta = 192^{\circ}  \to 360^{\circ} $
for the inverted mass ordering (IO) at 3$\sigma$ CL \cite{Esteban:2020cvm}.
The mixing matrix has two large mixings, which is very different from the quark mixing.
Apart from the tiny masses of neutrinos, 
such flavor structures will give us hints of physics beyond the SM.

One candidate behind the lepton sector is non-Abelian discrete flavor symmetries, such as $S_{3}$, $A_{4}$ and $S_{4}$ (see Refs. \cite{Altarelli:2010gt,Ishimori:2010au,King:2013eh,King:2014nza,King:2017guk,Perez:2019aqq} for reviews).\footnote{Applications of modular symmetries to explain the neutrino flavor structure have been proposed 
(see {\it e.g.}, Ref.\cite{Feruglio:2017spp}), where the Yukawa couplings are restricted by the modular symmetry.
}
In particular, the study of the $A_4$ models has received considerable interest.
It has been shown in Ref. \cite{Altarelli:2005yx} that the $A_4$ flavor symmetry leads naturally to 
the neutrino mass matrix that gives  tri-bimaximal flavor mixing,  $M_{\rm Tri}$
 ({\it i.e.}, $s_{12} = 1/\sqrt{3} ,~s_{23} = 1\sqrt{2},~s_{13} = 0$) \cite{Harrison:2002er}.
It is known that $M_{\rm Tri}$ is given by
a linear combination of three flavor structures as
\begin{align}
	M_{{\rm Tri}} = \frac{m_{1} + m_{3}}{2}
\begin{pmatrix}
1 & 0 & 0\\0 & 1 & 0\\0 & 0 & 1
\end{pmatrix}
+\frac{m_{2} - m_{1}}{3}
\begin{pmatrix}
1 & 1 & 1\\1 & 1 & 1\\1 & 1 & 1
\end{pmatrix}
+\frac{m_{1} - m_{3}}{2}
\begin{pmatrix}
1 & 0 & 0\\0 & 0 & 1\\0 & 1 & 0
\end{pmatrix}.
\end{align}
However, since the observed value of $\theta_{13}$ is small but non-zero, the neutrino mass matrix
should be modified from $M_{{\rm Tri}}$
so as to realize the non-zero (1,3) off-diagonal element  
in the flavor mixing matrix. 
One possible form of the neutrino mass matrix that derives the non-zero $\theta_{13}$ is given by
adding another new flavor structure as \cite{Shimizu:2011xg}
\begin{align}
	M_{\nu} = a
	\begin{pmatrix}	1 & 0 & 0\\		0 & 1 & 0\\		0 & 0 & 1		\end{pmatrix}	+b
	\begin{pmatrix}	1 & 1 & 1\\	1 & 1 & 1\\	1 & 1 & 1	\end{pmatrix}	+c
	\begin{pmatrix}	1 & 0 & 0\\	0 & 0 & 1\\	0 & 1 & 0	\end{pmatrix}	+d
	\begin{pmatrix}	0 & 0 & 1\\	0 & 1 & 0\\	1 & 0 & 0	\end{pmatrix},
	\label{mnu}
\end{align}
where the coefficients of each flavor structure $a,~ b,~ c$ and $d$ are the arbitrary-mass dimensionful parameters.
The non-vanishing $d$ term in the models with  $A_4$ symmetry is discussed in Refs. 
 \cite{Ishimori:2010au,Shimizu:2011xg,Sierra:2014hea,Karmakar:2016cvb,Novichkov:2018yse,Kobayashi:2019xvz,Okada:2020rjb}.
It is expected that the relationship between these four flavor structures 
will provide us with important information on the flavor symmetry in the neutrino mass generation mechanism.

In this paper, we propose a hybrid seesaw model based on the non-Abelian $A_{4}$ flavor symmetry, in which the neutrino mass matrix Eq.~({\ref{mnu}}) is generated 
by the tree-level and the one-loop seesaw mechanisms.\footnote{Other hybrid seesaw models based on the $A_{4}$ flavor symmetry have been considered in Refs. \cite{Chen:2005jm,Borah:2013lva,Sierra:2014hea,Borah:2014fga,Hamada:2014xha,Mukherjee:2015axj,Pramanick:2015qga,Franco:2015pva,Pramanick:2017fdq,Borah:2017dmk,Laamara:2018zpo,Mishra:2019keq,Wang:2019xbo}.}
These mechanisms generate different flavor structures, 
which leads to a characteristic hierarchy between the coefficients of the four flavor structures.
The $d$ term in Eq.~({\ref{mnu}}) comes from the one-loop seesaw mechanism. 
Two benchmark points are chosen in our model and
their predictions for the effective neutrino mass and the Majorana CP phases are computed.
Before presenting the results in our model, we also show the predictions of a 
model-independent analysis by using Eq.~({\ref{mnu}}) for some cases with hierarchical flavor structure.
The $A_4$ symmetry in our model is broken into the $Z_2$ subgroup by the vacuum expectation value (VEV) of the $A_4$ triplet scalar field.
Therefore, the lightest neutral $Z_2$-odd field, where we assume it to be a CP-even neutral scalar field, is stable and becomes a DM candidate. 
We compute the relic abundance and the spin-independent 
cross section of the DM, and show the plausible mass region of the DM.

\section{Model}
The non-Abelian $A_{4}$ flavor symmetry has four irreducible representations which are three singlets ${\bf 1},~{\bf 1'}$ and ${\bf 1''}$ and one triplet ${\bf 3}$.
The  $A_{4}$ symmetry is generated by two elements $S$ and $T$,
\begin{align*}
	S =
	\begin{pmatrix}
		1 & 0 & 0\\
		0 & -1 & 0\\
		0 & 0 & -1
	\end{pmatrix},\qquad
	T = \begin{pmatrix}
		0 & 1 & 0\\
		0 & 0 & 1\\
		1 & 0 & 0
	\end{pmatrix},
\end{align*}
which fulfill the relations $S^{2} = T^{3} = (ST)^{3} = I$.
The $A_{4}$ triplets $3_a=(a_1,a_2,a_3)$ and $3_b=(b_1, b_2, b_3)$ have the following multiplication rules:
\begin{align}
	\left[ 3_a\otimes 3_b\right]_{1} 
&= a_{1}b_{1} + a_{2}b_{2} + a_{3}b_{3}, \nonumber\\
\left[3_a\otimes 3_b\right]_{1'} 
&= a_{1}b_{1} + \omega a_{2}b_{2} + \omega^{2}a_{3}b_{3}, \nonumber\\
\left[3_a\otimes 3_b\right]_{1''} 
&= a_{1}b_{1} + \omega^{2}a_{2}b_{2} + \omega a_{3}b_{3}, \nonumber\\
\left[3_a\otimes 3_b\right]_{3_{1}} &= ( a_{2}b_{3},  a_{3}b_{1} , a_{1}b_{2} ), \nonumber\\
\left[3_a\otimes 3_b\right]_{3_{2}} &= ( a_{3}b_{2},  a_{1}b_{3} , a_{2}b_{1} )\nonumber ,
\end{align}
where $\omega = e^{\frac{2\pi}{3}i}$ which satisfies $1 + \omega + \omega^{2} = 0$.

\begin{table}[t]
\begin{center}
\begin{tabular}{|c|c|c|c|}\hline
 & $A_{4}$ & $SU(2)$ & $Z_{2}$ \\\hline
 $(L_{e},L_{\mu},L_{\tau})$ & $1,1'',1'$ & $2$ & $(+,+,+)$ \\\hline
 $(l_{e_{R}},l_{\mu_{R}},l_{\tau_{R}})
 $ & $1,1',1''$ & $1$ & $(+,+,+)$ \\\hline
 $(N_{1},N_{2},N_{3})$ & $3$ & $1$ & $(+,-,-)$ \\\hline
$H$ & $1$ & $2$  & $+$\\\hline
$\eta = (\eta_{1},\eta_{2},\eta_{3})$ & $3$ & $2$ & $(+,-,-)$ \\\hline
\end{tabular}
\end{center}
\caption{$A_{4}$ flavor and $SU(2)$ gauge quantum numbers for leptons, right-handed neutrinos and scalar fields of the model.}\label{table1}
\end{table}

 We introduce three $A_4$ triplet right-handed neutrinos $N_{i}$  ($i = 1, 2, 3$), which are invariant under the SM gauge group,
 and three $A_4$ triplet $SU(2)$ scalar doublets $\eta_{i}$. Model assignments are shown in Table \ref{table1}, where the  $A_{4}$ singlets $L_{e},~L_{\mu},~L_{\tau}$ ($l_{e_{R}},~l_{\mu_{R}},~l_{\tau_{R}}$) are lepton doublets (lepton singlets) and the $A_{4}$ singlet $H$ is a Higgs doublet field.
In this model, the Yukawa sectors of neutrinos are described by
\begin{align}
	\mathcal{L}_{{\rm Yukawa}} &= y_{1}\overline{L_{e}}(\tilde \eta_{1} N_{1}+ \tilde \eta_{2} N_2+\tilde \eta_{3}N_3)
	+ y_{1'}\overline {L_{\mu}}(\tilde \eta_{1} N_1 + \omega \tilde \eta_{2} N_2+ \omega^{2}\tilde \eta_{3}N_3)\nonumber\\ 
	&+ y_{1''}\overline{L_{\tau}}(\tilde \eta_{1} N_1+ \omega^{2} \tilde \eta_{2} N_2+ \omega \tilde \eta_{3}N_3)+\rm{h.c.},
\label{yukawa}
\end{align}
where $y_{1},~y_{1'}$ and $y_{1''}$ are the Yukawa couplings and $\tilde \eta_i =i\sigma^2 \eta_{i}^\ast$. 
In this work we assume $y \equiv y_{1} = y_{1'} = y_{1''}$ for simplicity,
 \footnote{The exact alignment of the Yukawa coupling ($y_1 = y_{1'} = y_{1''}$) is not necessary to explain the oscillation parameters. The deviations of $y_{1'}$ and $ y_{1''}$ from $y_{1}$ are respectively within about 7 \% and 3 \% for explaining the observed neutrino oscillation parameters at 3$\sigma$ CL \cite{Esteban:2020cvm}.}
where $y$ is real.
Majorana mass terms of right-handed neutrinos are given by
\begin{align}
	\mathcal{L}_{{\rm Majorana}} &= M_{R}(\overline {N_{1}^c}N_{1} + \overline {N_{2}^c} N_{2} + \overline {N_{3}^c}N_{3}) +M'_{R}(\overline {N_{1}^c}N_{1} + \omega \overline {N_{2}^c}N_{2} +\omega^{2} \overline {N_{3}^c}N_{3})\nonumber\\
& +M''_{R}(\overline {N_{1}^c}N_{1} + \omega^{2} \overline {N_{2}^c}N_{2} +\omega \overline {N_{3}^c}N_{3}) + M_{23}(\overline{N_{2}^c}N_{3} + {\rm h.c.}),
\label{rneumass}
\end{align}
where $M_{R},~ M_{R}',~ M_{R}''$ and $M_{23}$ are the Majorana masses of right-handed neutrinos.
We note that the second, third, and fourth therms in Eq.~(\ref{rneumass}) break the $A_{4}$ symmetry. \footnote {These three terms can be generated by $A_{4}$ singlet ${\bf 1''}$, ${\bf 1'}$ and triplet ${\bf 3}$ scalar fields, respectively ( the generation of the fourth term is discussed in, {\it e.g.}, Ref. \cite{delaVega:2018cnx}).}
Because of the fourth term, the neutrinos $N_2$ and $N_3$ are mixed with each other. 
The mass matrix of right-handed neutrinos is diagonalized by the complex mixing angle $\tan2\hat{\theta}_{R}\equiv \frac{2M_{23}}{(\omega - \omega^{2})(M_{R}'' - M_{R}')}$, where we define the diagonal elements as $(M_{1},M_{2}e^{i\delta_{R_{2}}},M_{3}e^{i\delta_{R_{3}}})$.
Throughout this paper, we work in the basis where the charged lepton mass matrix is diagonal.

Based on the $A_{4}$ symmetry, the scalar potential is given by \cite{Boucenna:2011tj}
\begin{align}
V &= \mu_{\eta}^{2}[\eta^{\dagger}\eta]_{1} + \mu_{H}^{2}H^{\dagger}H + \lambda_{1}(H^{\dagger}H)^{2} + \lambda_{2}[\eta^{\dagger}\eta]_{1}^{2} + \lambda_{3}[\eta^{\dagger}\eta]_{1'}[\eta^{\dagger}\eta]_{1''}\nonumber\\
&+ \lambda_{4}[\eta^{\dagger}\eta^{\dagger}]_{1'}[\eta\eta]_{1''} + \lambda_{4}'[\eta^{\dagger}\eta^{\dagger}]_{1''}[\eta\eta]_{1'} + \lambda_{5}[\eta^{\dagger}\eta^{\dagger}]_{1}[\eta\eta]_{1} + \lambda_{6}\left( [\eta^{\dagger}\eta]_{3_{1}}[\eta^{\dagger}\eta]_{3_{1}} + {\rm h.c.} \right)\nonumber\\
&+ \lambda_{7}[\eta^{\dagger}\eta]_{3_{1}}[\eta^{\dagger}\eta]_{3_{2}} + \lambda_{8}[\eta^{\dagger}\eta^{\dagger}]_{3_{1}}[\eta\eta]_{3_{2}} + \lambda_{9}[\eta^{\dagger}\eta]_{1}[H^{\dagger}H] + \lambda_{10}[\eta^{\dagger}H]_{3_{1}}[H^{\dagger}\eta]_{3_{1}}\nonumber\\
&+ \lambda_{11}\left( [\eta^{\dagger}\eta^{\dagger}]_{1}HH + {\rm h.c.} \right) + \lambda_{12}\left( [\eta^{\dagger}\eta^{\dagger}]_{3_{1}}[\eta H]_{3_{1}} + {\rm h.c.} \right)
+ \lambda_{13}\left( [\eta^{\dagger}\eta]_{3_{1}}[ \eta^{\dagger}H ]_{3_{1}} + {\rm h.c.}\right) \nonumber\\
&+ \lambda_{14}\left( [\eta^{\dagger}\eta]_{3_{2}}[\eta^{\dagger}H]_{3_{1}} +{\rm h.c.}\right).
 \label{eq:V}
\end{align}
We assume that the couplings in the scalar potential are real and $\lambda_{4} = \lambda_{4}'$ for simplicity.
We note that this assumption means requiring CP conservation in the scalar potential.
 When one of the $A_{4}$ triplet field $\eta_{1}$, in addition to $H$, has the VEV, the $A_{4}$ symmetry breaks to the subgroup $Z_{2}$ symmetry 
 whose charge assignments are also shown in Table \ref{table1}. The $Z_{2}$-even fields $H$ and $\eta_{1}$ are defined as
\begin{align}
H =
\begin{pmatrix}
\phi^{+}\\
\frac{1}{\sqrt 2}(v_{h} + \phi^0 + i\chi)
\end{pmatrix}
,\;\;\;
\eta_{1} =
\begin{pmatrix}
w_{1}^{+}\\
\frac{1}{\sqrt 2}(v_{\eta} + \eta_{1_R} + i\eta_{1_I})
\end{pmatrix}.
\end{align}
Here the VEVs are real and satisfy $\sqrt{v_{h}^{2} + v_{\eta}^{2}} = v = 246 ~{\rm GeV}$.
The stationary conditions are given by
\begin{align}
&\mu_{H}^{2} + \lambda_{1}v_{h}^{2} + \frac{1}{2}(\lambda_{9} + \lambda_{10} + 2\lambda_{11})v_{\eta}^{2} = 0,\\
&\mu_{\eta}^{2} + (\lambda_{2} + \lambda_{3} + 2\lambda_{4} + \lambda_{5})v_{\eta}^{2} + \frac{1}{2}(\lambda_{9} + \lambda_{10} + 2\lambda_{11})v_{h}^{2} = 0.
\end{align}
The physical scalar states in the $Z_2$-even sector can be obtained  by the mixing angles $\beta$, where $\tan\beta \equiv v_{\eta}/v_{h}$,
 and $\alpha$ as
\begin{align}
\begin{pmatrix}G^+\\H^+\end{pmatrix}
 &=\begin{pmatrix}\cos\beta &\sin\beta\\-\sin\beta & \cos\beta\end{pmatrix}
\begin{pmatrix}\phi^+ \\\omega_{1}^+\end{pmatrix},~~~~
\begin{pmatrix}G^0\\A_{1}\end{pmatrix}
 =\begin{pmatrix}\cos\beta &\sin\beta\\-\sin\beta & \cos\beta\end{pmatrix}
\begin{pmatrix}\chi \\\eta_{_{1I}}\end{pmatrix},~~~
\end{align}
\begin{align}
\begin{pmatrix} h_2\\h_1 \end{pmatrix}
 &=
\begin{pmatrix}\cos\alpha &\sin\alpha\\-\sin\alpha & \cos\alpha\end{pmatrix}
\begin{pmatrix}	\phi^{0} \\\eta_{_{1R}}\end{pmatrix}.~~~~
\end{align}
Here $h_1$ is the SM-like Higgs particle whose mass is $m_{h_1}=125$ GeV. 
The masses of the charged scalar field $H^{\pm}$ and the CP-odd scalar field $A_{1}$ are described as $m^{2}_{H^{\pm}} = -\frac{1}{2}(\lambda_{10} + 2\lambda_{11})v^{2}$ and $m^{2}_{A_{1}} = -2\lambda_{11}v^{2}$, respectively.

The $Z_{2}$-odd fields $\eta_{2}$ and $\eta_{3}$, which do not have VEVs, are defined as
\begin{align}
\eta_{2} &=
\begin{pmatrix}\eta_{2}^{+}\\ \frac{1}{\sqrt{2}}( \eta_{_{2R}} + i\eta_{_{2I}} )
\end{pmatrix}
,\;\;\;
\eta_{3}=
\begin{pmatrix}
\eta_{3}^{+}\\
\frac{1}{\sqrt{2}}( \eta_{_{3R}} + i\eta_{_{3I}} )
\end{pmatrix}.
\end{align}
These two states are mixed through the $\lambda_{12},~\lambda_{13}$, and $\lambda_{14}$ terms in Eq.~(\ref{eq:V}) 
and the mixing angle between them is $\pi/4$.
The neutral CP-even (-odd) states 
give the mass eigenstates $\eta_{2}^0$ and $\eta_{3}^0$  ($A_{2}$ and $A_{3}$)
with masses
$m_{\eta_{2}^0}$ and $m_{\eta_{3}^0}$  ($m_{A_{2}}$ and $m_{A_{3}}$) as
\begin{align}
	m_{\eta_{2}^{0}}^{2} &= \frac{1}{2}\left(
	 \lambda_{x_{1}}v_{\eta}^{2}  - 3\lambda_{x_{3}}v_{\eta}v_{h} \right) ,\\
	 m_{\eta_{3}^{0}}^{2} &= \frac{1}{2}\left(
	 \lambda_{x_{1}}v_{\eta}^{2} + 3\lambda_{_{x_{3}}}v_{\eta}v_{h}\right) ,\\
	 m_{A_{2}}^{2} &= \frac{1}{2}\left(
	\lambda_{x_{2}}v_{\eta}^{2} -4 \lambda_{11}v_{h}^{2} - \lambda_{x_{3}}v_{\eta}v_{h} \right) ,\\
	m_{A_{3}}^{2} &= \frac{1}{2}\left(
	\lambda_{x_{2}}v_{\eta}^{2} -4 \lambda_{11}v_{h}^{2}  + \lambda_{x_{3}}v_{\eta}v_{h}\right) ,
\end{align}
where $  
  \lambda_{x_{1}} \equiv -3\lambda_{3} - 6\lambda_{4} + 2\lambda_{6} + \lambda_{7} + \lambda_{8} ,~
  \lambda_{x_{2}} \equiv -3\lambda_{3} - 2\lambda_{4} - 4\lambda_{5} - 2\lambda_{6} + \lambda_{7} + \lambda_{8}$ and $
  \lambda_{x_{3}} \equiv \lambda_{12} + \lambda_{13} + \lambda_{14}$.
The masses of the
 charged scalar fields $\eta_{2}^{\pm}$ and $\eta_{3}^{\pm}$ are given by
\begin{align}
	m_{\eta^{\pm}_{2}}^{2} &= \frac{1}{2}\left[ \lambda_{x_{4}}v_{\eta}^{2} + (\lambda_{10} + \lambda_{11})v_{h}^{2} - \lambda_{x_{3}}v_{h}v_{\eta}\right],\\
	m_{\eta^{\pm}_{3}}^{2} &= \frac{1}{2}\left[ \lambda_{x_{4}}v_{\eta}^{2} + (\lambda_{10} + \lambda_{11})v_{h}^{2} + \lambda_{x_{3}}v_{h}v_{\eta}\right],
\end{align}
where $
  \lambda_{x_{4}} \equiv -3\lambda_{3} - 4\lambda_{4} - 2\lambda_{5} + \lambda_{8}$.
 Note that the mass differences between $m_{\eta_{2}^{0}}$ and $m_{\eta_{3}^{0}}$, $m_{A_{2}}$ and $m_{A_{3}}$, $m_{\eta_{2}^{\pm}}$ and $m_{\eta_{3}^{\pm}}$ are given by $\lambda_{x_{3}}$.

 The lightest $Z_{2}$-odd particle is stable and can be a DM if it is neutral. In our model, the right-handed neutrinos are heavy as shown later, so that the DM candidates are $\eta_{2,3}^{0}$ and $A_{2,3}$. Hereafter, we assume that $\eta_{2}^{0}$ is a DM candidate.

\section{Neutrino masses and flavor structures}
\begin{figure}[b]
\centering
\includegraphics[keepaspectratio,scale=0.6]{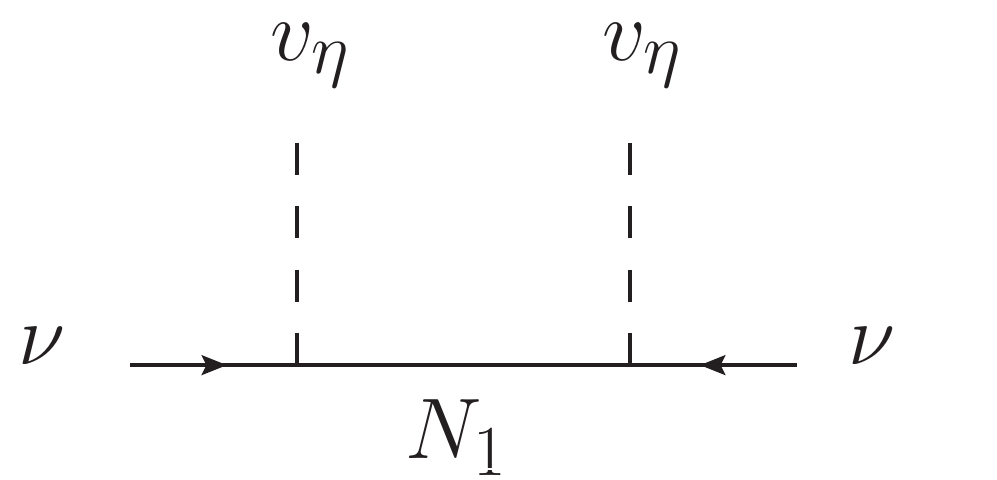}
\caption{Feynman diagram for neutrino mass via the tree-level seesaw mechanism.}
\label{fig:type-i}
\end{figure}

In this model, the neutrinos obtain their masses via the tree-level and the one-loop seesaw mechanisms.
Since the Yukawa interactions $L_\alpha\tilde HN_i (\alpha=e, \mu, \tau)$ are forbidden by the $A_4$ symmetry, the usual Type-I seesaw mechanism does not work.
However, the neutrinos can obtain their masses via the other tree-level seesaw mechanism due to the existence of the Yukawa
interactions $ L_\alpha \tilde\eta_1N_1$ in Eq.~(\ref{yukawa}) with the non-zero VEV of $\eta_1$.  
The neutrino mass matrix $M_{\nu}^{{\rm tree}}$ that is generated by the tree-level seesaw mechanism in Fig.~\ref{fig:type-i} is given by
\begin{align}
	M_{\nu}^{{\rm tree}}  
	=\frac{v_{\eta}^{2}}{2M_{1}} 
	\begin{pmatrix}		y& 0 & 0\\		y & 0 & 0\\		y& 0 & 0	\end{pmatrix}
		\begin{pmatrix}		y &y&y \\		0 & 0 & 0\\		0 & 0 & 0	\end{pmatrix}
 = \frac{v_{\eta}^{2}y^{2}}{2M_{1}}
\begin{pmatrix}1 & 1 & 1\\1 & 1 & 1\\1 & 1 & 1\end{pmatrix}.
	\label{Type-I}
\end{align}
The flavor structure in $M_{\nu}^{{\rm tree}}$ is the same form as that of the $b$ term in Eq.~(\ref{mnu}).\footnote{It is noted that when the $A_{4}$ symmetry is broken to the $Z_{3}$ symmetry via $\braket{\eta} = (v_\eta,v_\eta,v_\eta,)$, the $b,~ c$, and $d$ terms are generated by the tree-level seesaw mechanism.}
We note that the rank of $M_{\nu}^{{\rm tree}}$ is one, so that other contributions to
the neutrino mass generation should be necessary.

\begin{figure}[t]
\begin{minipage}{0.45\hsize}
\centering
\includegraphics[keepaspectratio,scale=0.6]{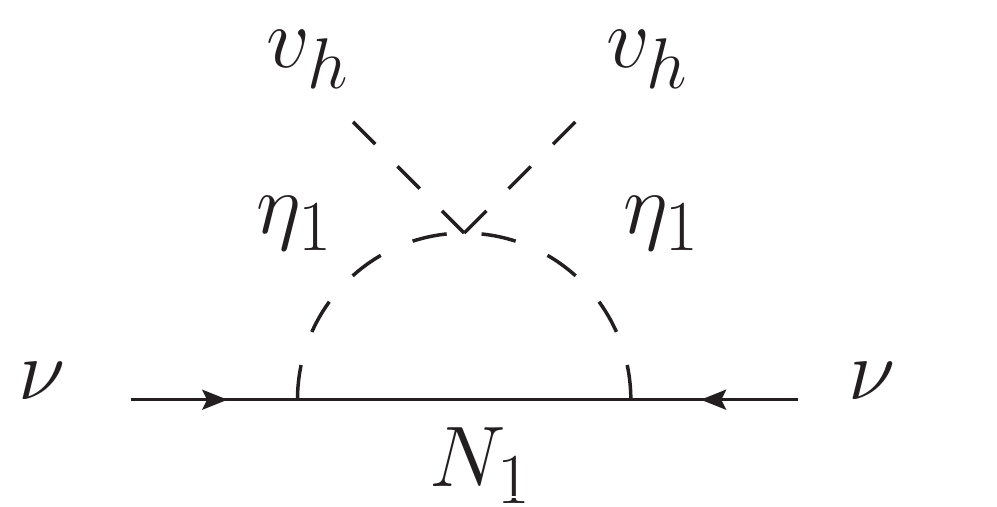}
\end{minipage}
\begin{minipage}{0.45\hsize}
\centering
\includegraphics[keepaspectratio,scale=0.6]{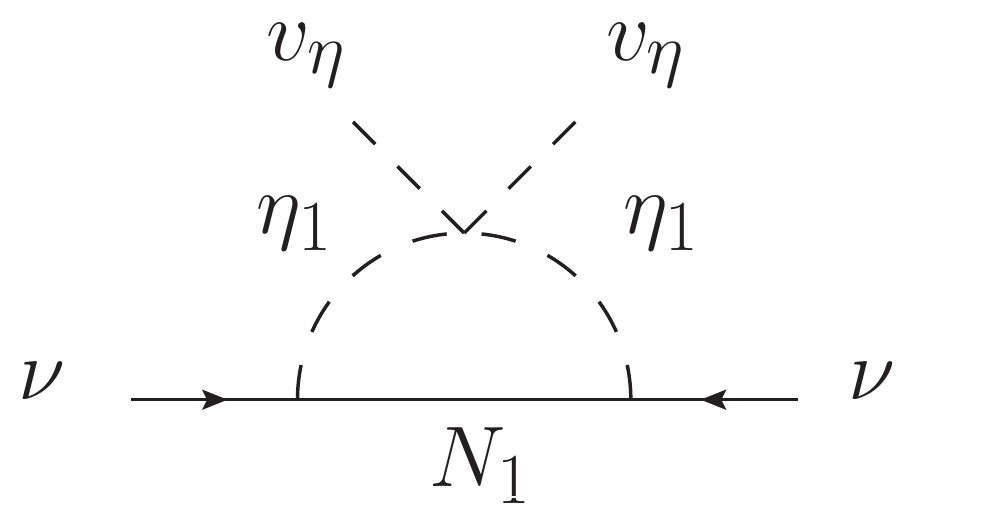}
\end{minipage}
\caption{Feynman diagrams for neutrino masses via the one-loop seesaw mechanism with $N_{1}$.}
\label{fig:one-r1}
\end{figure}

\begin{figure}[t]
\begin{minipage}{0.45\hsize}
\centering
\includegraphics[keepaspectratio,scale=0.6]{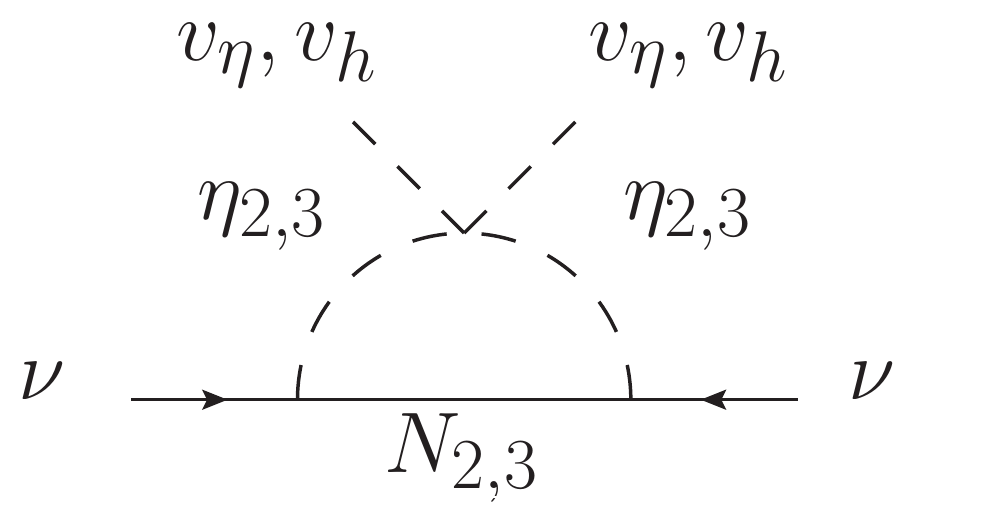}
\end{minipage}
\begin{minipage}{0.45\hsize}
\centering
\includegraphics[keepaspectratio,scale=0.6]{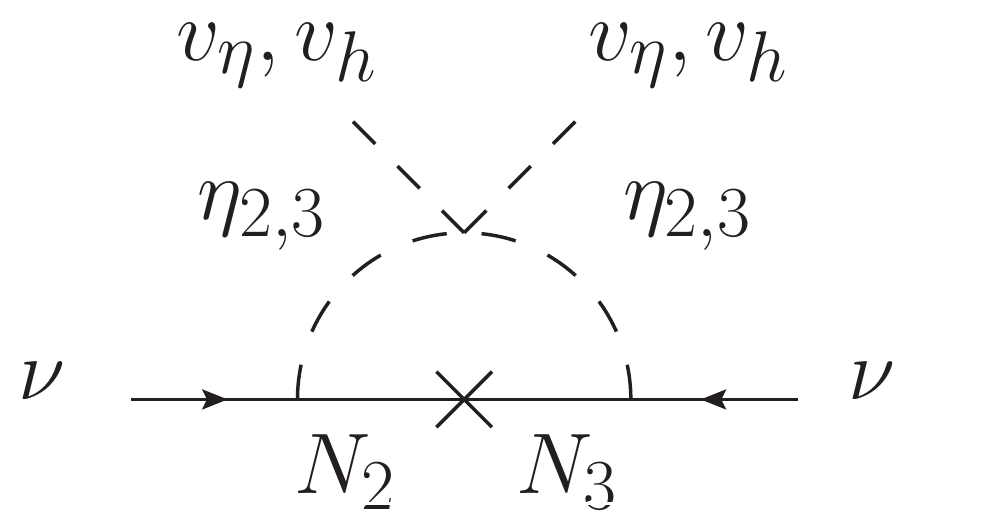}
\end{minipage}
\caption{Feynman diagrams for neutrino masses via the one-loop seesaw mechanism  with $N_{2}$ and $N_{3}$.}
\label{fig:one-r23}
\end{figure}

The neutrino masses are also generated by the one-loop seesaw mechanisms that are shown in Figs.~\ref{fig:one-r1} and \ref{fig:one-r23}.
The $Z_{2}$-even right-handed neutrino $N_{1}$ contributes to the mass generation in Fig.~\ref{fig:one-r1},
while the $Z_{2}$-odd right-handed neutrinos 
$N_{2}$, $N_{3}$ and their mixing contribute in Fig.~\ref{fig:one-r23}.
Assuming $\lambda_{x_{3}} \ll 1$ and $\sin(\beta - \alpha) = 1$, where the former assumption leads to 
$m_{\eta} \equiv m_{\eta_{2}^{0}} \approx m_{\eta_{3}^{0}}$ and $m_{A} \equiv m_{A_{2}} \approx m_{A_{3}}$,
the neutrino mass matrix via the one-loop diagrams in Figs.~\ref{fig:one-r1} and \ref{fig:one-r23} is described as
\begin{align}
M^{{\rm one\mathchar`-loop}} &= 
 \begin{pmatrix}
		1 & 1 & 1\\
		1 & \omega & \omega^{2}\\
		1 & \omega^{2} & \omega
	\end{pmatrix}
\begin{pmatrix}
1 & 0 & 0\\
0 & \cos\hat{\theta}_{R}  & \sin\hat{\theta}_{R}\\
0 & -\sin\hat{\theta}_{R}  & \cos\hat{\theta}_{R}\\
\end{pmatrix}
\begin{pmatrix}
\Lambda_{1} & 0 & 0\\
0 & \Lambda_{2} & 0\\
0 & 0 &\Lambda_{3}
\end{pmatrix}
\begin{pmatrix}
1 & 0 & 0\\
0 & \cos\hat{\theta}_{R}  & -\sin\hat{\theta}_{R}\\
0 & \sin\hat{\theta}_{R}  & \cos\hat{\theta}_{R}\\
\end{pmatrix}
 \begin{pmatrix}
		1 & 1 & 1\\
		1 & \omega & \omega^{2}\\
		1 & \omega^{2} & \omega
	\end{pmatrix}\nonumber\\
&= X_{a}
\begin{pmatrix}
1 & 0 & 0\\
0 & 1 & 0\\
0 & 0 & 1
  \end{pmatrix}
+\left[ \Lambda_{1} - \frac{X_{a} + X_{c} + X_{d}}{3} \right]
\begin{pmatrix}
1 & 1 & 1\\
1 & 1 & 1\\
1 & 1 & 1
  \end{pmatrix}
+ X_{c}
\begin{pmatrix}
1 & 0 & 0\\
0 & 0 & 1\\
0 & 1 & 0
  \end{pmatrix}
+ X_{d}
\begin{pmatrix}
0 & 0 & 1\\
0 & 1 & 0\\
1 & 0 & 0
  \end{pmatrix}.
\label{nmassone}
\end{align}
Here 
\begin{align}
\Lambda_{1} &\equiv \frac{y^{2}M_{1}}{16\pi^{2}}\left[\sin^{2}\beta\frac{m_{h_{1}}^{2}}{m_{h_{1}}^{2} - M_{1}^{2}}\ln\frac{m_{h_{1}}^{2}}{M_{1}^{2}} + \cos^{2}\beta\left(\frac{m_{h_{2}}^{2}}{m_{h_{2}}^{2} - M_{1}^{2}}\ln\frac{m_{h_{2}}^{2}}{M_{1}^{2}} - \frac{m_{A_{1}}^{2}}{m_{A_{1}}^{2} - M_{1}^{2}}\ln\frac{m_{A_{1}}^{2}}{M_{1}^{2}}\right)\right],\\
\Lambda_{k} &\equiv \frac{y^{2}M_{k}e^{\delta_{R_{k}}}}{16\pi^{2}}\left[\frac{m_{\eta}^{2}}{m_{\eta}^{2} - M_{k}^{2}}\ln\frac{m_{\eta}^{2}}{M_{k}^{2}} - \frac{m_{A}^{2}}{m_{A}^{2} - M_{k}^{2}}\ln\frac{m_{A}^{2}}{M_{k}^{2}}\right]~~~~(k=2,3), \\
X_{a} &\equiv 3\cos\hat{\theta}_{R}\sin\hat{\theta}_{R}(\Lambda_{3} - \Lambda_{2}),\\
X_{c} &\equiv \left[ (1-\omega) (\Lambda_{2}\cos^{2}\hat{\theta}_{R} + \Lambda_{3}\sin^{2}\hat{\theta}_{R}) + (1 - \omega^{2}) (\Lambda_{2}\sin^{2}\hat{\theta}_{R} + \Lambda_{3}\cos^{2}\hat{\theta}_{R})\right],\\
X_{d} &\equiv \left[(\omega^{2}-\omega) (\Lambda_{2}\cos^{2}\hat{\theta}_{R} + \Lambda_{3}\sin^{2}\hat{\theta}_{R}) + (\omega - \omega^{2}) (\Lambda_{2}\sin^{2}\hat{\theta}_{R} + \Lambda_{3}\cos^{2}\hat{\theta}_{R})\right].
\end{align}
It can be seen that the four flavor structures in Eq.~(\ref{mnu})
are generated. 
\footnote{Even when the assumption $\lambda_{x_{3}} \ll 1$ is removed, the four flavor structures  are derived.} 
We find that the one-loop diagrams with $N_{1}$ in Fig.~\ref{fig:one-r1} generate only the $b$ term.
On the other hand, the contributions of $N_{2}$ and $N_{3}$ in Fig.~\ref{fig:one-r23} give all four flavor structures.
In particular, the mixing between $N_{2}$ and $N_{3}$ realizes the non-zero $a$ term, while
the origin of the non-zero $d$ term ({\it i.e.}, non-zero $\theta_{13}$) comes from the difference between $\Lambda_{2}$ and $\Lambda_{3}$.
From Eqs.~(\ref{Type-I}) and (\ref{nmassone}), the neutrino mass matrix in our model is given by
\begin{align}
M_{\nu} &= M_{\nu}^{{\rm tree}} + M^{{\rm one\mathchar`-loop}} \nonumber\\
&= X_{a}
\begin{pmatrix}
1 & 0 & 0\\
0 & 1 & 0\\
0 & 0 & 1
  \end{pmatrix}
+ \left[ \frac{v_{\eta}^{2}y^{2}}{2M_{1}} + \Lambda_{1} - \frac{X_{a} + X_{c} + X_{d}}{3} \right]
\begin{pmatrix}
1 & 1 & 1\\
1 & 1 & 1\\
1 & 1 & 1
  \end{pmatrix}
+ X_{c}
\begin{pmatrix}
1 & 0 & 0\\
0 & 0 & 1\\
0 & 1 & 0
  \end{pmatrix}
+ X_{d}
\begin{pmatrix}
0 & 0 & 1\\
0 & 1 & 0\\
1 & 0 & 0
  \end{pmatrix},
\label{nmassmodel}
\end{align}
where the flavor structures are the same as those in Eq.~(\ref{mnu}).
In this model, the ratios $|b|/|c|$ and $|b|/|d|$ are naively given by the inverse of the loop suppression factor $\approx 16\pi^{2}$ when 
$M_{i} \sim m_{\eta}$, since the $b$ term contains the contributes from the tree-level diagrams whereas the $c$ and $d$ terms are generated by the one-loop diagrams.
However, such large hierarchies between $b$ and $c,~ d$ are not plausible with the current experimental data,
as will be shown later. 
On the other hand, when $M_{i} \gg m_{\eta}$, 
we obtain $M_{\nu}^{{\rm tree}}/\Lambda_{i} \propto 16\pi^{2}\left[ \ln\frac{M_{i}^{2}}{m_{\eta}^{2}} - 1 \right]^{-1}$.
As the mass difference between $M_{i}$ and $m_{\eta}$ becomes larger, the ratio $M_{\nu}^{{\rm tree}}/\Lambda_{i}$ becomes smaller.
Therefore, milder (but large) hierarchies between $b$ and $c,~ d$, such as $|b|/|c|\sim |b|/|d| \sim \mathcal{O}(10)$, can be possible.
For the $a$ term, 
although it is also generated by the one-loop diagrams, 
its magnitude is controlled by the mixing between $N_2$ and $N_3$.


The neutrino mass matrix in Eq.~(\ref{mnu}) (and thus in Eq.~(\ref{nmassmodel})) is diagonalized 
with the PMNS matrix that is formed by the tri-bimaximal mixing, the (1,3) mixing and  the Majorana phase matrix
 \cite{Shimizu:2011xg}:
\begin{align}
U_{{\rm PMNS}} = 
\begin{pmatrix}
\frac{2}{\sqrt{6}} & \frac{1}{\sqrt{3}} & 0\\
-\frac{1}{\sqrt{6}} & \frac{1}{\sqrt{3}} & -\frac{1}{\sqrt{2}}\\
-\frac{1}{\sqrt{6}} & \frac{1}{\sqrt{3}} & \frac{1}{\sqrt{2}}\\
\end{pmatrix}
\begin{pmatrix}
\cos\hat{ \theta}  & 0 & \sin\hat{ \theta}\\
0 & 1 & 0\\
-\sin\hat{ \theta}  & 0 & \cos\hat{ \theta}
\end{pmatrix} 
	\begin{pmatrix}
		1 & 0 & 0\\
		0 & e^{i\alpha_{2}/2} & 0\\
		0 & 0 & e^{i\alpha_{3}/2}
			\end{pmatrix}.
\end{align}
Here the mixing angle $\hat{ \theta} $ is the complex parameter.
Comparing this to the standard parametrization of $U_{{\rm PMNS}}$ in Eq.~(\ref{UPMNS}), we obtain
$\sin\hat{ \theta} =\sqrt{3/2}\sin\theta_{13} e^{-i\delta}$. 
Using the coefficient parameters $a,~b,~c$, and $d$ in Eq.~(\ref{mnu}), the angle $\hat{ \theta}$ is given  
by
\begin{align}
	\tan 2\hat{ \theta} = \frac{\sqrt{3}d}{-2c + d}\,.
	\label{theta_hat}
\end{align}
The neutrino masses $m_{1},~m_{2}$, and $m_{3}$ and
the Majorana phases $\alpha_{2}$ and  $\alpha_{3}$ are written as
\begin{align}
	m_{1} &= \left| a + \sqrt{c^{2} + d^{2} - cd} \right|,~~~
m_{2} = \left|a + 3b + c + d\right|,~~~
m_{3} = \left|a - \sqrt{c^{2} + d^{2} - cd}\right|, \label{m1}\\
	\alpha_{2} &= {\rm arg}(a + 3b + c + d) - {\rm arg}\left(a + \sqrt{c^{2} + d^{2} - cd} \right),~~~  \\
 	\alpha_{3} &= {\rm arg}\left(a - \sqrt{c^{2} + d^{2} - cd}\right) - {\rm arg}\left(a + \sqrt{c^{2} + d^{2} - cd} \right).  
	\label{alpha3}
\end{align}
The observed values of the neutrino oscillation parameters $\Delta m_{31}^{2}$, $\Delta m_{21}^{2}$, $\theta_{13}$ and $\delta$ give  constraints on the relationship between the coefficients $a,~b,~c$, and $d$.

\begin{figure}[t]
\centering
\includegraphics[keepaspectratio,scale=0.3]{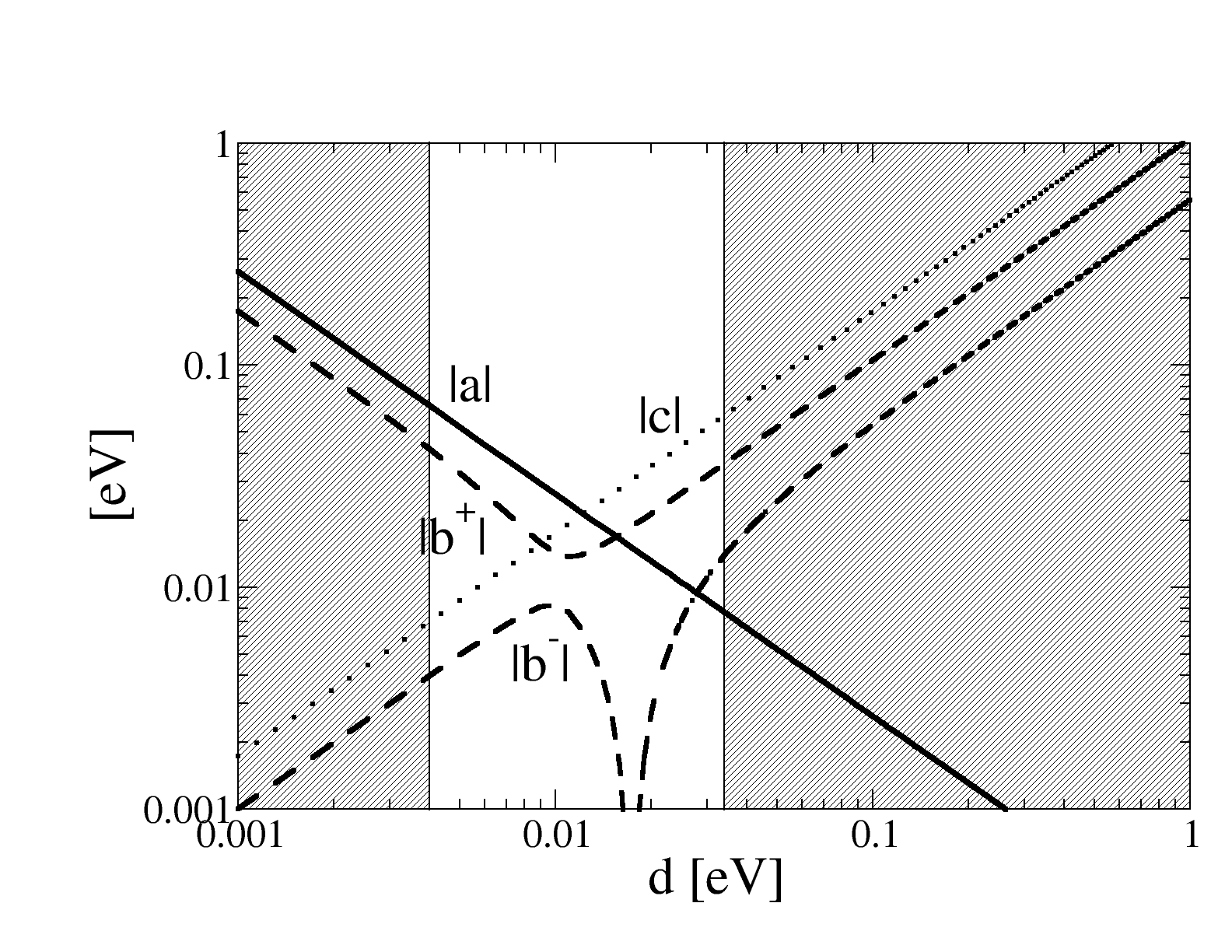}~~~~~
\includegraphics[keepaspectratio,scale=0.3]{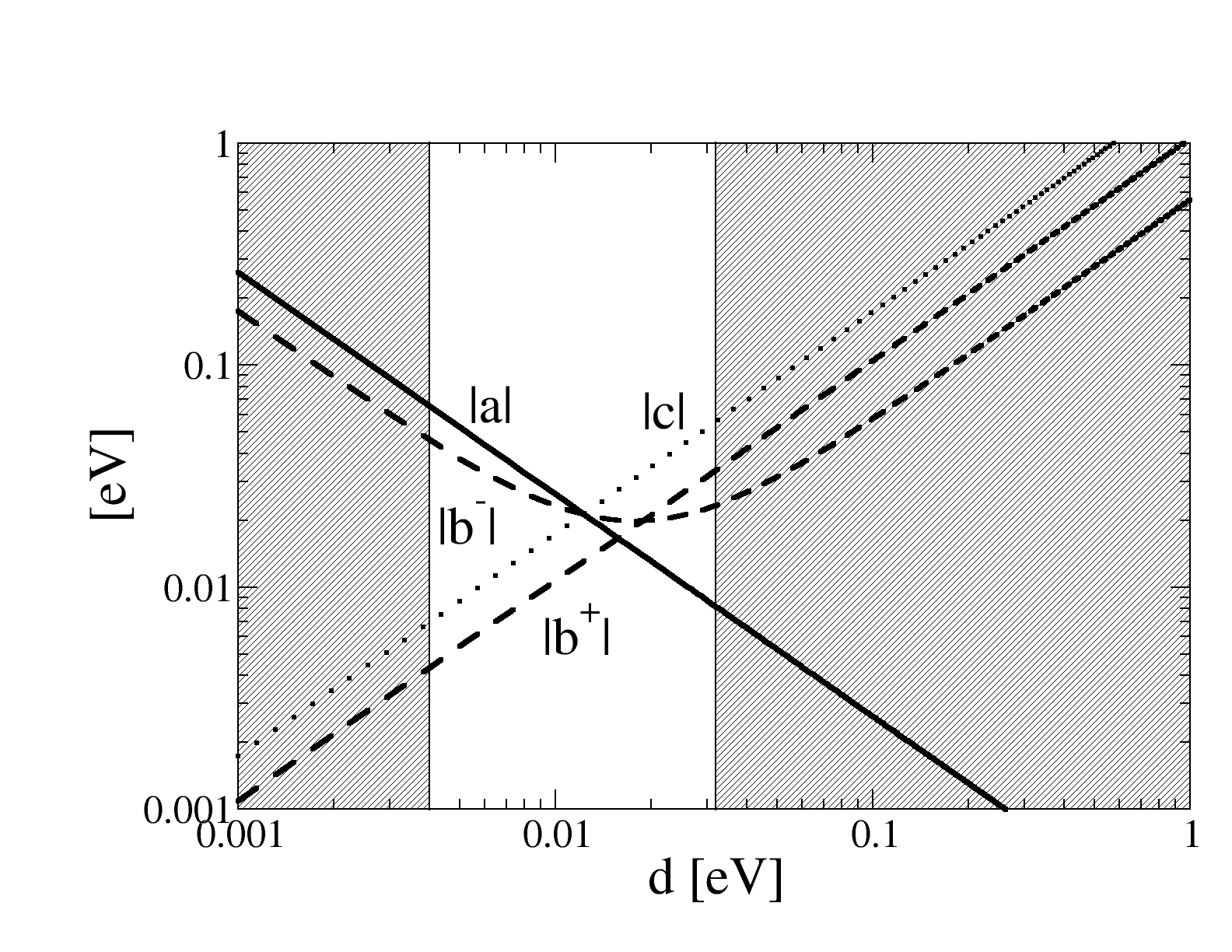}
\caption{
The coefficients of the flavor structure $|a|$, $|b|$, and $|c|$ as a function of $d$ for the NO (left panel) and the IO (right panel).
The values of the neutrino oscillation parameters are taken 
as in  Eq.~(\ref{center}) with $\delta = 0$.
The shaded regions are excluded by the constraint of the sum of light neutrino masses \cite{Aghanim:2018eyx}.
}
\label{fig:abcd}
\end{figure}

In Fig.~\ref{fig:abcd},  we show 
the absolute values of the coefficients $|a|$ (solid lines), $|b|$ (dashed lines) and $|c|$ (dotted lines) 
as a function of $d$ that is assumed to be real and positive, for the NO (left panel) and the IO (right panel).
The coefficients are derived from the following center values in the NO (IO) \cite{Esteban:2020cvm}:\footnote{We use the center values in v4.1 of Ref.~\cite{Esteban:2020cvm}.}
\begin{align}
\Delta m_{21}^{2} = 7.39\times 10^{-5} \;{\rm eV}^{2},~~~
\Delta m_{31}^{2} = 2.525~&(-2.512)\times 10^{-3} \;{\rm eV}^{2},~~~
\theta_{13} = 8.61~(8.65)^{\circ}.
\label{center}
\end{align}
The Dirac phase is taken as $\delta = 0$ and 
the coefficients $a,~ b$, and $c$ are assumed to be real for simplicity.
There are two solutions for $|b|$, 
which are shown by $|b^+|$ and $|b^-|$ in Fig.~\ref{fig:abcd}.
The shaded regions 
are excluded by the constraint of the sum of light neutrino masses using the Planck TT, TE, EE+lowE+lensing \cite{Aghanim:2018eyx}:
\begin{eqnarray}
	\sum_{i} m_{i} < 0.241 ~{\rm eV}  
	\label{meeconst}.
\end{eqnarray}
In the NO (left panel of Fig.~\ref{fig:abcd}),  
  $|a|$ decreases and $|c|$ increases as the parameter $d$ becomes larger.
We note that the coefficients $|c|$ and $|d|$ are comparable but $|c|$ is larger than $|d|$ due to the relation of Eq.~(\ref{theta_hat}). 
The hierarchy $|a|, |b^+| > |c|,|d|$ is shown for the smaller $d$  ($d\lesssim 0.008~{\rm eV}$), while
 the hierarchy $ |b^+|, |c|, |d| > |a|$ appears for the larger $d$ ($d\gtrsim 0.02~{\rm eV}$).
 On the other hand, the hierarchy $|a| ,|c|,|d| > |b^-|$ is obtained for $d\sim 0.03~{\rm eV}$, where $b^{-}$ changes its sign.
In the IO (right panel), similar hierarchies among the coefficients can be seen except for the disappearance of $|b^-|$.\footnote{In the left panel of Fig.~\ref{fig:abcd}, 
 $b^+$ is positive ({\it i.e.}, ${\rm arg}(b^+) = 0$),
 $b^-$ is positive for $d < 0.02~{\rm eV}$ 
and negative ({\it i.e.}, ${\rm arg}(b^-) = \pi~{\rm rad}$) 
 for $d > 0.02~{\rm eV} $, and $a$ and $c$ are negative. 
In the right panel of Fig.~\ref{fig:abcd},  $a$, $b^+$, $b^-$, and $c$ are positive, positive, negative, and negative, respectively. 
}
We note that, in both the NO and the IO, the large hierarchies between $b$ and $c$, $d$ such as
$|b|/|c|$, $|b|/|d|$ $\approx 16\pi^{2}$ are disfavored by current data.

In our hybrid seesaw model, the milder but large hierarchy of $|b|/|d| \approx \pi^{2}$ can be realized for $M_{i} \gg m_{\eta}$ as mentioned in the previous section.
In the following, we give the benchmark point in the NO (BP$_{{\rm NO}}$) and the IO (BP$_{{\rm IO}}$), respectively, 
where the ratio $|b|/|d| \approx \pi^{2}$ is satisfied.
Here we take the coefficients $a,~ b,~ c$, and $d$ as complex parameters and take into account the contributions of CP phases.
For BP$_{{\rm NO}}$, we take the following set for $a,~ b,~ c$, and $d$, 
which satisfies the center values of the neutrino parameters in Eq.~(\ref{center}) and $\delta = 222^{\circ}$ \cite{Esteban:2020cvm}:
\begin{align}
&|a| \approx 0.0759 ~{\rm eV},~~~|b| \approx 0.0483 ~{\rm eV} ,~~~|c| \approx 0.0103 ~{\rm eV} ,~~~|d| \approx 0.0045 ~{\rm eV},\nonumber\\
&{\rm arg}(a) \approx 2.45~{{\rm rad}},~~~{\rm arg}(b) \approx -0.439~{{\rm rad}},~~~{\rm arg}(c) \approx 2.09~{{\rm rad}},~~~{\rm arg}(d) \approx 1.44~{{\rm rad}}
\label{bench_abcd}
\end{align}
The above set is realized by the following values of the model parameters:  
\begin{align}
&\tan\beta = 3,~~y= 1.0\times 10^{-2} ,~~ (M_{1},~M_{2},~M_{3}) \approx (6.69,~1.94,~1.53) \times 10^{10}~{\rm GeV} ~~\nonumber\\ &
 \delta_{R_{2}} \approx -0.66 ~{\rm rad},~~\delta_{R_{3}} \approx 2.43 ~{\rm rad},~~ \tan2\hat{\theta}_{R} \approx -16.5 + 10.4i, \nonumber\\ &
m_{h_{2}} = 200~{\rm GeV},~~m_{A_{1}} = 250~{\rm GeV},~~ m_{\eta} = 500 ~{\rm GeV}, ~~
m_{A} = 520 ~{\rm GeV}.
\label{bench}
\end{align}
For the IO, we take the following set for the BP$_{{\rm IO}}$, which satisfies Eq.~(\ref{center}) and 
$\delta = 285^{\circ}$ \cite{Esteban:2020cvm}:
\begin{align}
&|a| \approx 0.0707 ~{\rm eV},~~~|b| \approx 0.0536 ~{\rm eV} ,~~~|c| \approx 0.00978 ~{\rm eV} ,~~~|d| \approx 0.0049 ~{\rm eV},\nonumber\\
&{\rm arg}(a) \approx 0.063~{{\rm rad}},~~~{\rm arg}(b) \approx 3.14~{{\rm rad}},~~~{\rm arg}(c) \approx 0.044~{{\rm rad}},~~~{\rm arg}(d) \approx -1.51~{{\rm rad}}
\label{bench_abcdIO}
\end{align}
The above set is realized by the following:
\begin{align}
&\tan\beta = 3,~~y = 1.0\times 10^{-2} ,~~(M_{1},~M_{2},~M_{3}) \approx (6.02,~2.16,~1.61) \times10^{10}~{\rm GeV}~~\nonumber\\&
 \delta_{R_{2}} \approx -3.12 ~{\rm rad},~~\delta_{R_{3}} \approx 0.091 ~{\rm rad},~~ \tan2\hat{\theta}_{R} \approx -16.7,\nonumber\\ &
m_{h_{2}} = 200~{\rm GeV},~~m_{A_{1}} = 250~{\rm GeV},~~ m_{\eta} = 500 ~{\rm GeV}, ~~
m_{A} = 520 ~{\rm GeV}.
\label{benchIO}
\end{align}
From Eqs.~(\ref{bench}) and (\ref{benchIO}), we find that the large hierarchy $|b|/|d| \approx \pi^{2}$ is realized for 
the masses of the right-handed neutrinos
 $M_{i} \sim \mathcal{O}(10^{10})~{\rm GeV}$ and the scalar fields $m_{\eta_{i}} \sim \mathcal{O}(10^{2{\rm \mathchar`-}3})~{\rm GeV}$ for $y \sim 10^{-2}$. 
Furthermore, both the BP$_{{\rm NO}}$ and the BP$_{{\rm IO}}$ satisfy $|a| \gg |c|,|d|$, so that the Majorana phase $\alpha_3$ is expected to be close to zero, as can be seen from Eq.~(\ref{alpha3}).
%
%
 \section{ Predictions of the effective neutrino mass and the Majorana phases }
In this section, we discuss the predictions of the effective neutrino mass $m_{ee}$ and the Majorana CP phases. 
 $m_{ee}$ is defined as
\begin{align}
m_{ee} = \left|\sum_{i=1}^3 U_{ei}^{2}m_{i} \right|
\end{align}
with
$U_{e1} = 2\cos\hat{ \theta }/\sqrt{6},~U_{e2} = 1/\sqrt{3}$, and $
	U_{e3} = 2\sin\hat{ \theta }/\sqrt{6}$.
\begin{figure}[t]
	\centering
			\includegraphics[keepaspectratio,scale=0.28]{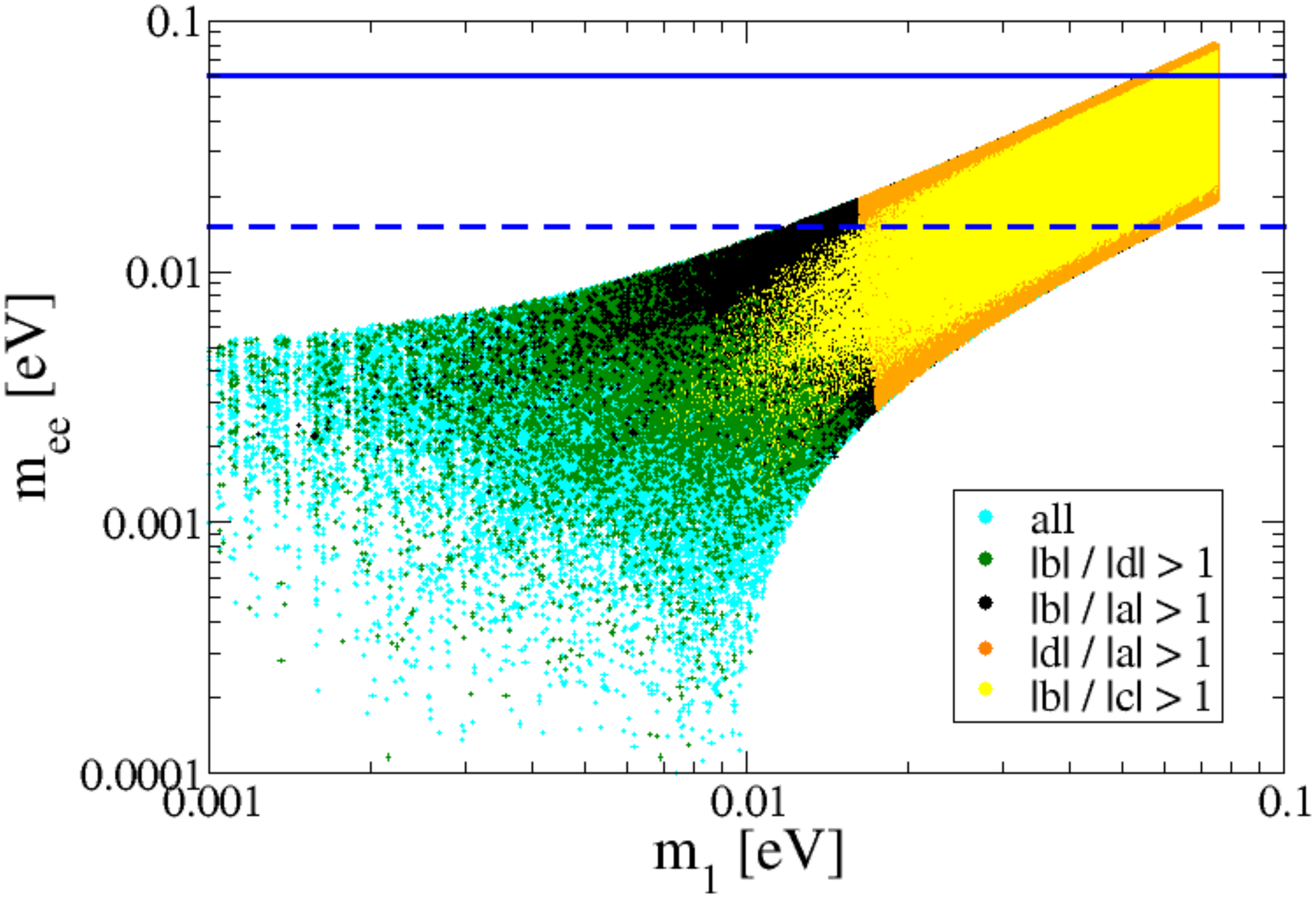}\\
	\includegraphics[keepaspectratio,scale=0.28]{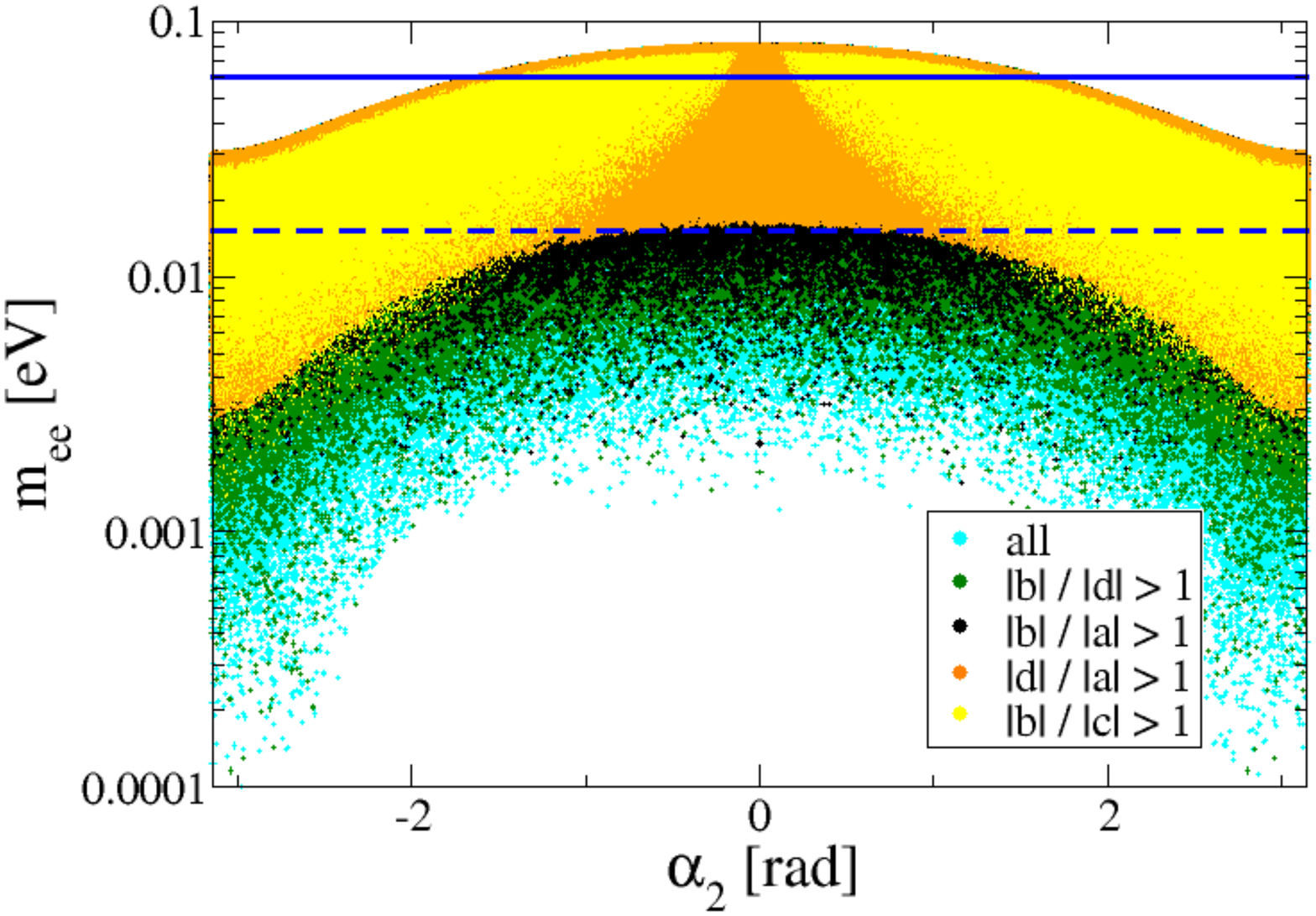}
		\includegraphics[keepaspectratio,scale=0.28]{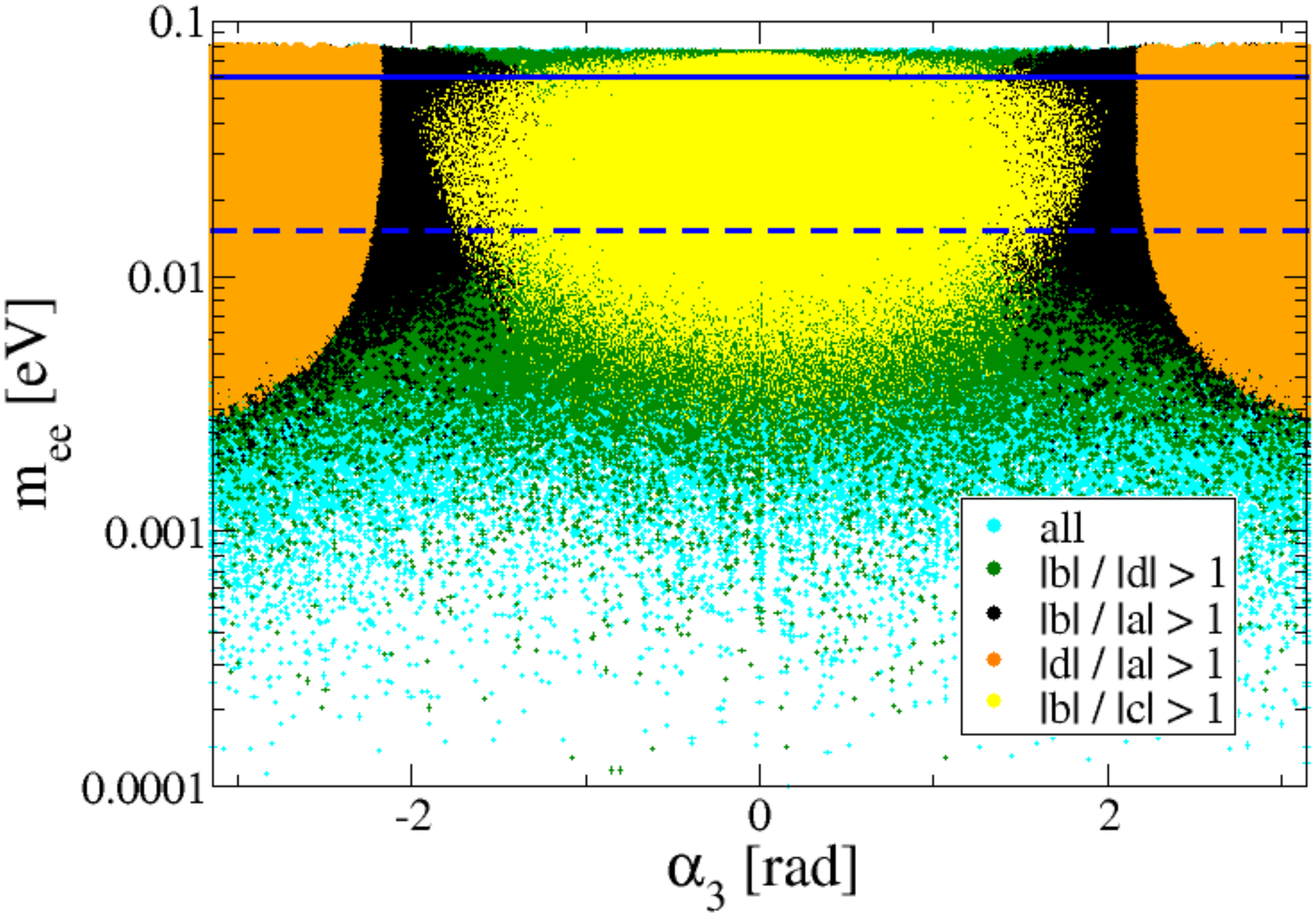}
\caption{$m_{ee}$ versus $m_1$ (upper panel),  $\alpha_2$ (lower left panel), $\alpha_3$ (lower right panel) for the NO.
The cyan points show all points that satisfy Eqs.~(\ref{center}) and (\ref{meeconst}),
the green, black, orange, and yellow points show respectively the hierarchy case with $|b| / |d| > 1$, $|b| / |a| > 1$, $|d| / |a| > 1$, and $|b| / |c| > 1$.
}
	\label{fig:nofighie}
\end{figure}
\begin{figure}[t]
	\centering
			\includegraphics[keepaspectratio,scale=0.28]{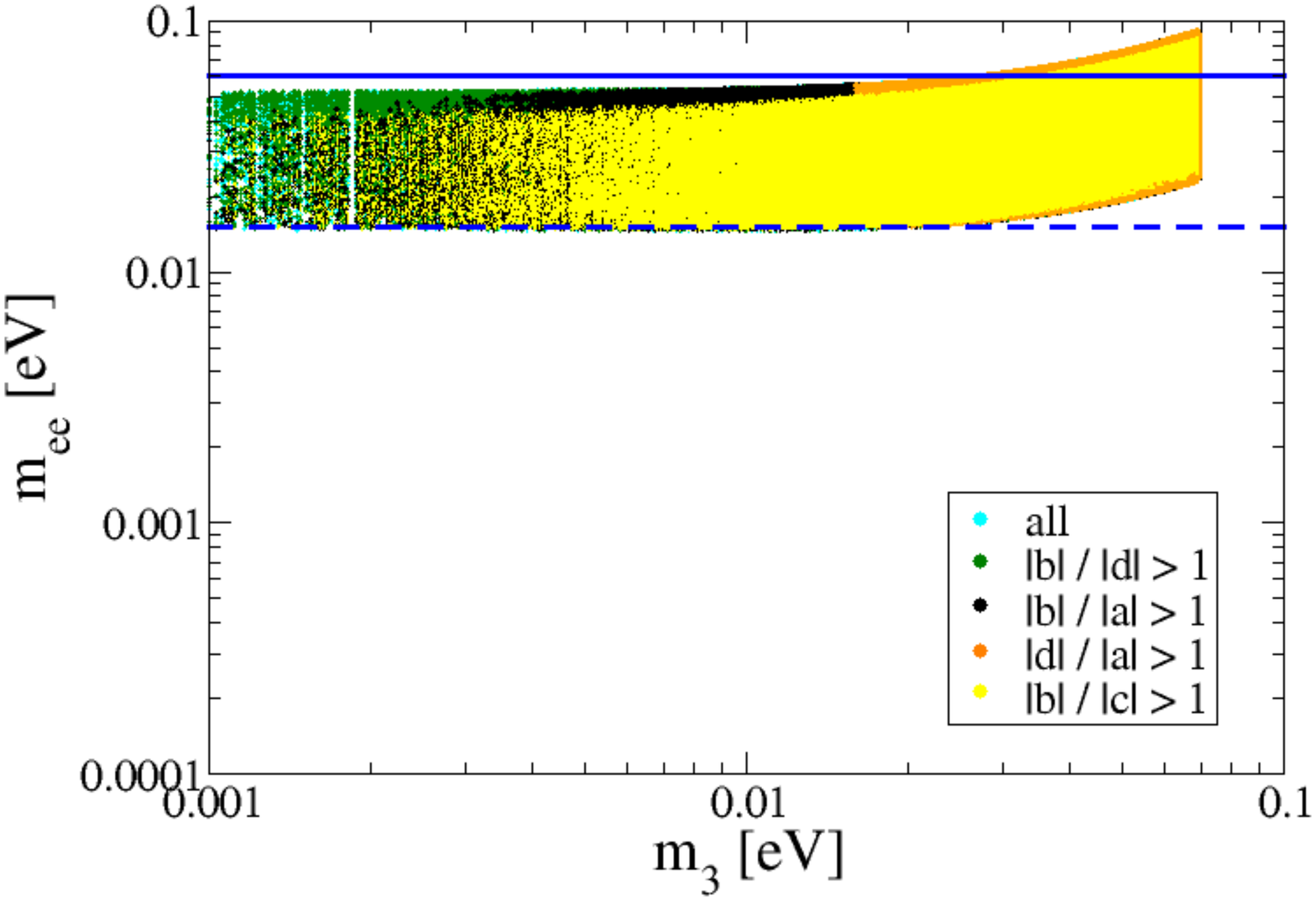}\\
			\includegraphics[keepaspectratio,scale=0.28]{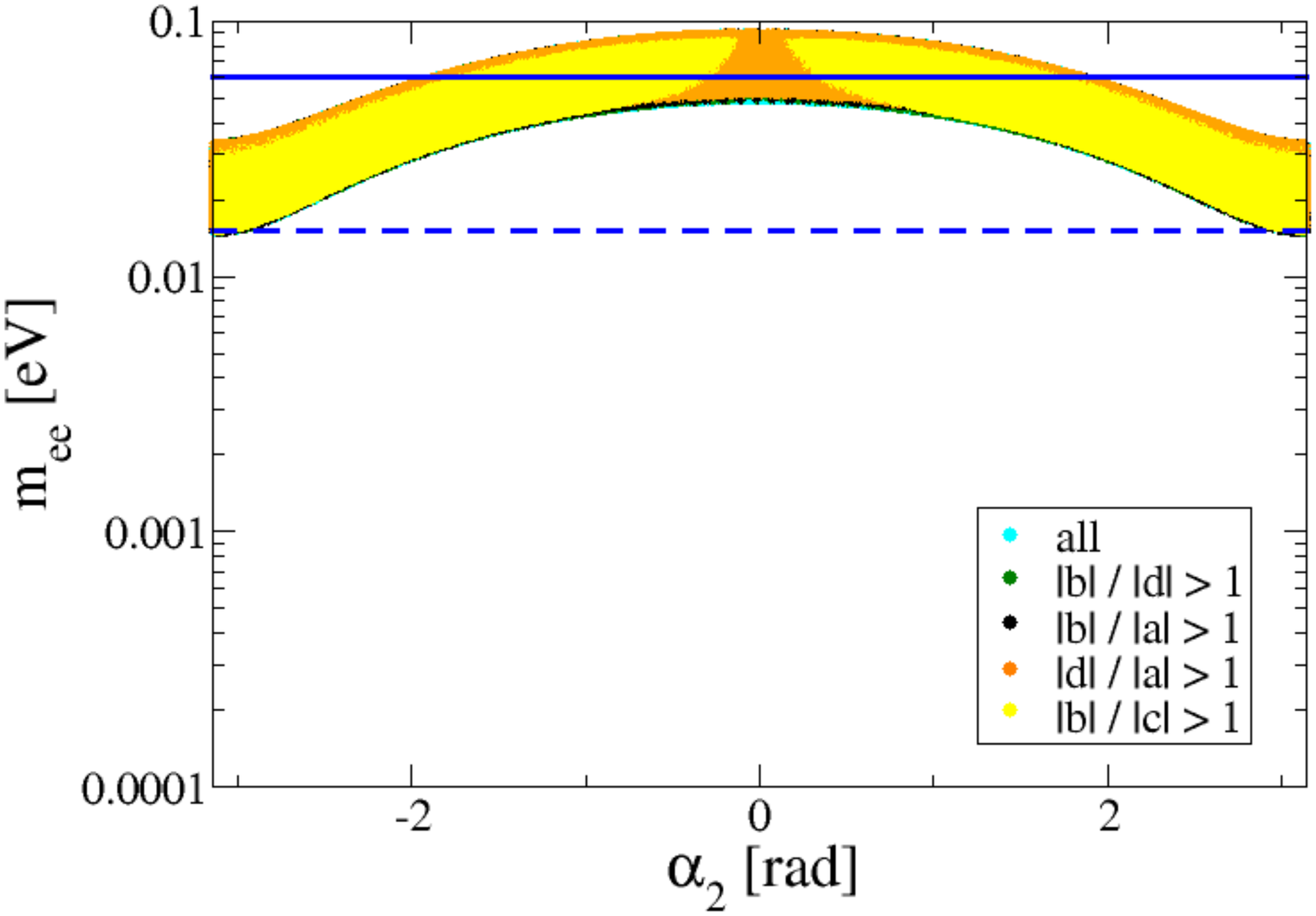}
			\includegraphics[keepaspectratio,scale=0.28]{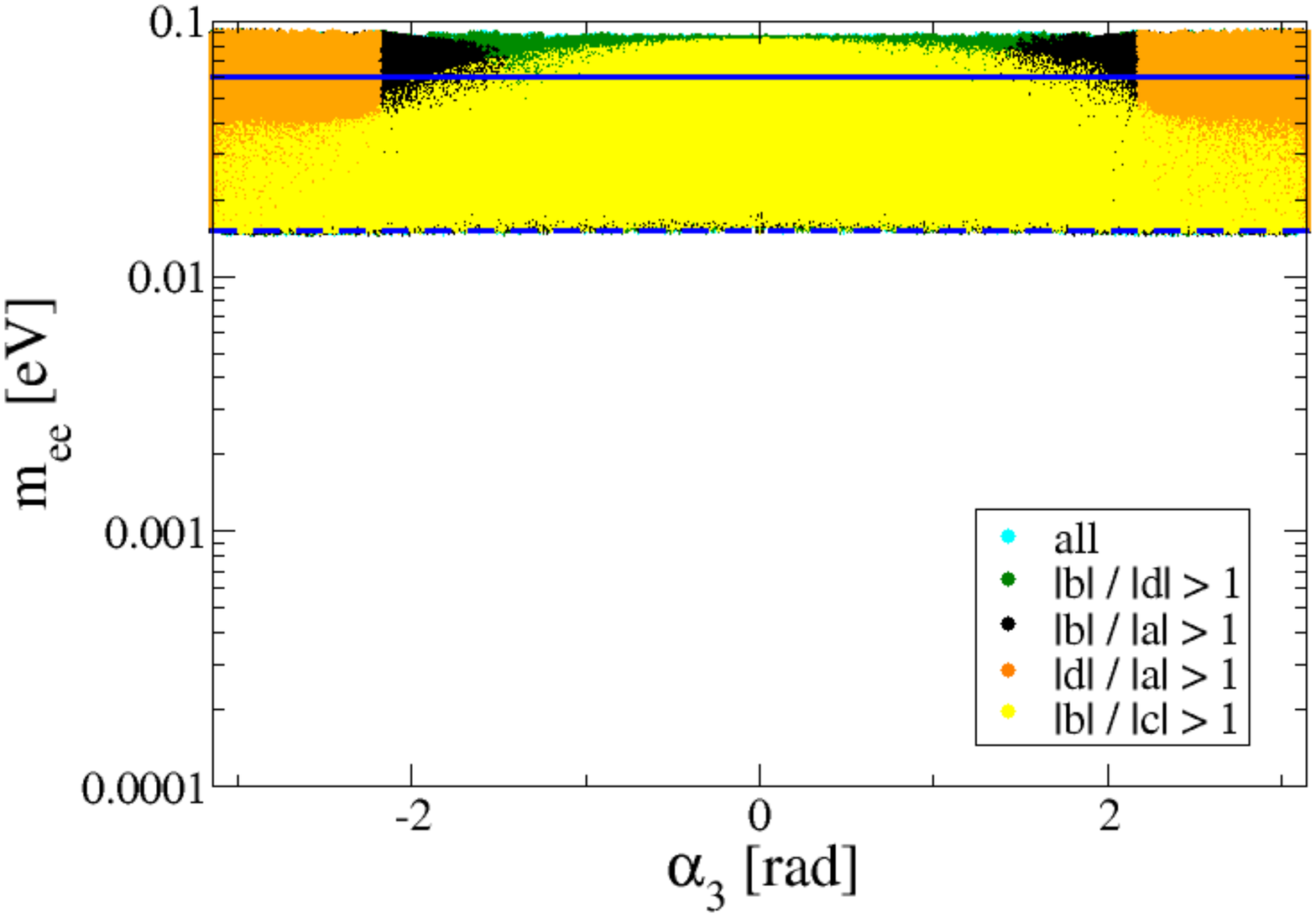}
\caption{Same as Fig.~\ref{fig:nofighie} except for the IO and the lightest neutrino mass $m_{3}$.}
	\label{fig:iofighie}
\end{figure}
First, 
we show the results of the model-independent analysis 
by using the neutrino mass matrix in Eq.~(\ref{mnu}),
focusing on the hierarchies between the coefficients $a$, $b$, $c$, and $d$.
In Figs.~\ref{fig:nofighie} and \ref{fig:iofighie},
the predicted values of $m_{ee}$ are shown as functions of the lightest neutrino mass (upper panel) and  
the Majorana phases $\alpha_2$ (lower left panel) and $\alpha_3$ (lower right panel) for the NO and the IO, respectively.
In these plots, we have taken the following ranges for the coefficient parameters $|d|$, ${\rm arg}(a)$, ${\rm arg}(b)$, ${\rm arg}(d)$, and the $3\sigma$ range of $\delta$ in the NO (IO) \cite{Esteban:2020cvm}:
\begin{align}
	&0 \leq |d|/{\rm eV} \leq 1.0 ~,~ 0 \leq {\rm arg}(a)/{\rm rad} < 2\pi~,~ 0\leq {\rm arg}(b)/{\rm rad} < 2\pi~,\nonumber\\
	&0 \leq {\rm arg}(d)/{\rm rad} < 2\pi~,~  107^{\circ}\leq \delta \leq 403^{\circ}~(192^{\circ} \leq \delta \leq 360^{\circ}).
	\label{para}     
\end{align}
The other parameters $|a|, |b|, |c|$ can be determined so as to satisfy the 
observed values of $\Delta m_{21}^{2}$, $\Delta m_{31}^{2}$ and $\theta_{13}$ in Eq.~(\ref{center}).
Furthermore, ${\rm arg}(c)$ is fixed through the relation in Eq.~(\ref{theta_hat}).
In Figs.~\ref{fig:nofighie} and \ref{fig:iofighie},
the cyan points show all points that satisfy the observed values in Eq.~(\ref{center}) and the constraint from Eq.~(\ref{meeconst}).
The green, black, orange and yellow points show the result where the hierarchical conditions 
 $|b| / |d| > 1$, $|b| / |a| > 1$, $|d| / |a| > 1$, and $|b| / |c| > 1$, respectively, are further imposed.
The horizontal blue solid and dashed lines show the upper bound on $m_{ee}$ by
the global fit of neutrinoless double beta decay ($0\nu\beta\beta$) experiments
$m_{ee} \lesssim 0.06~{\rm eV}$ \cite{Esteban:2020cvm} and
the sensitivity of the next-generation $0\nu\beta\beta$ experiment by nEXO \cite{Albert:2017hjq}.

In Fig.~\ref{fig:nofighie}, 
the hierarchical case with $|b| / |d| > 1$ (green) does not constrain the parameter space compared to the cyan points, while 
the other three cases
constrain the parameter space.
For the hierarchical case with $|b| / |c| > 1$ (yellow), 
the predicted regions of  the effective neutrino mass $m_{ee}$ and
the lightest neutrino mass $m_{1}$ are $m_{ee}\gtrsim 0.001~{\rm eV}$ and
$m_{1} \gtrsim 0.007~{\rm eV}$.
In this case, the Majorana phase $\alpha_{2} \sim 0$ is excluded by the constraint from Eq.~(\ref{meeconst}) and
 $\alpha_{3}$ is constrained as $|\alpha_{3}|/{\rm rad } \lesssim 2.0$. 
The hierarchical cases with $|d| / |a| > 1$ (orange) and $|b| / |a| > 1$ (black) have similar predictions, 
but the former is more constrained, such as giving $m_{1} \gtrsim 0.015~{\rm eV}$
and $|\alpha_{3}|/{\rm rad } \gtrsim 2.2$. Here $|\alpha_{3}|$ $\simeq \pi$ radians are obtained for $|d| \gg |a|$ as can be seen from Eq.~(\ref{alpha3}).
Figure~\ref{fig:iofighie} shows the results for the IO.
For the hierarchical case with $|b|/|c| > 1$, the predicted regions for 
$m_3$ and 
$\alpha_3$
are wider than those in the NO.
In particular, $\alpha_{3}$ is not constrained for 
$0.015 \lesssim m_{ee}/{\rm eV}\lesssim 0.04$.
Similarly, in that range of $m_{ee}$,
 $\alpha_{3}$ is unconstrained 
for the case with $|b|/|a| > 1$ (although this is not shown in the lower right panel in Fig.~\ref{fig:iofighie} 
as it is behind the yellow region).
The predictions for the 
cases with $|d| / |a| > 1$ show similar features to those in the NO.
The next-generation $0\nu\beta\beta$ experiment nEXO 
can explore all predicted regions of $m_{ee}$ for the IO,
and thus there is the possibility of obtaining  
hints to the Majorana phases for some hierarchical cases.
In this analysis, 
the predicted values for $\theta_{23}$ and $\theta_{12}$ are $0.4 \lesssim \sin^{2}\theta_{23} \lesssim 0.6$ and $\sin^{2}\theta_{12} \approx 0.34$, 
respectively,
which are allowed within $3\sigma$ \cite{Esteban:2020cvm}.
We note that the predicted regions of the hierarchies with $|b|/|c| > 1$ and $|d|/|a| > 1$ are included in those with $ |b|/|d| > 1$ and $|c|/|a| > 1$, respectively, because of the relation $|c| > |d|$ obtained by Eq.~(\ref{theta_hat}). 

\begin{figure}[t]
	\centering
			\includegraphics[keepaspectratio,scale=0.28]{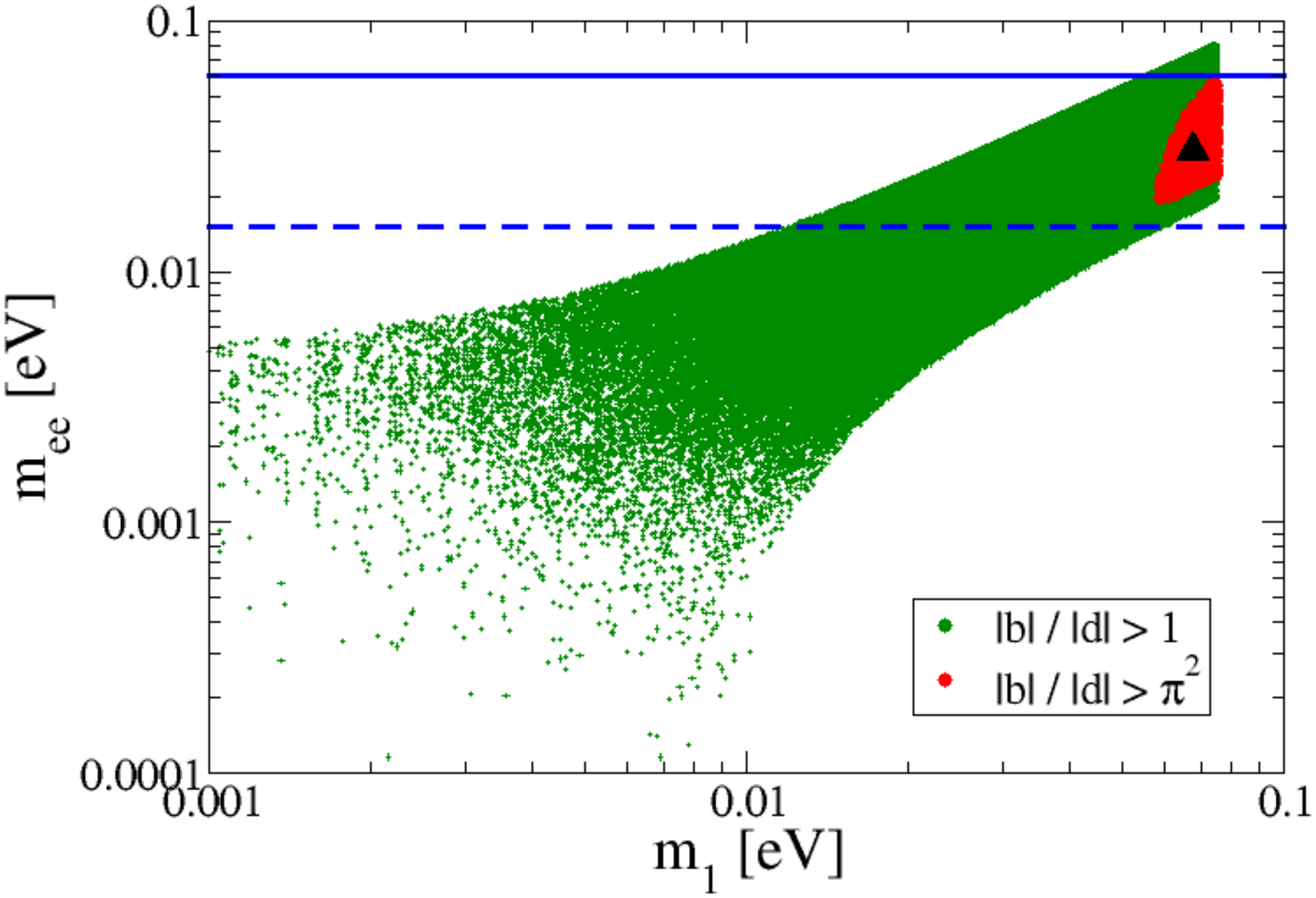}\\
			\includegraphics[keepaspectratio,scale=0.28]{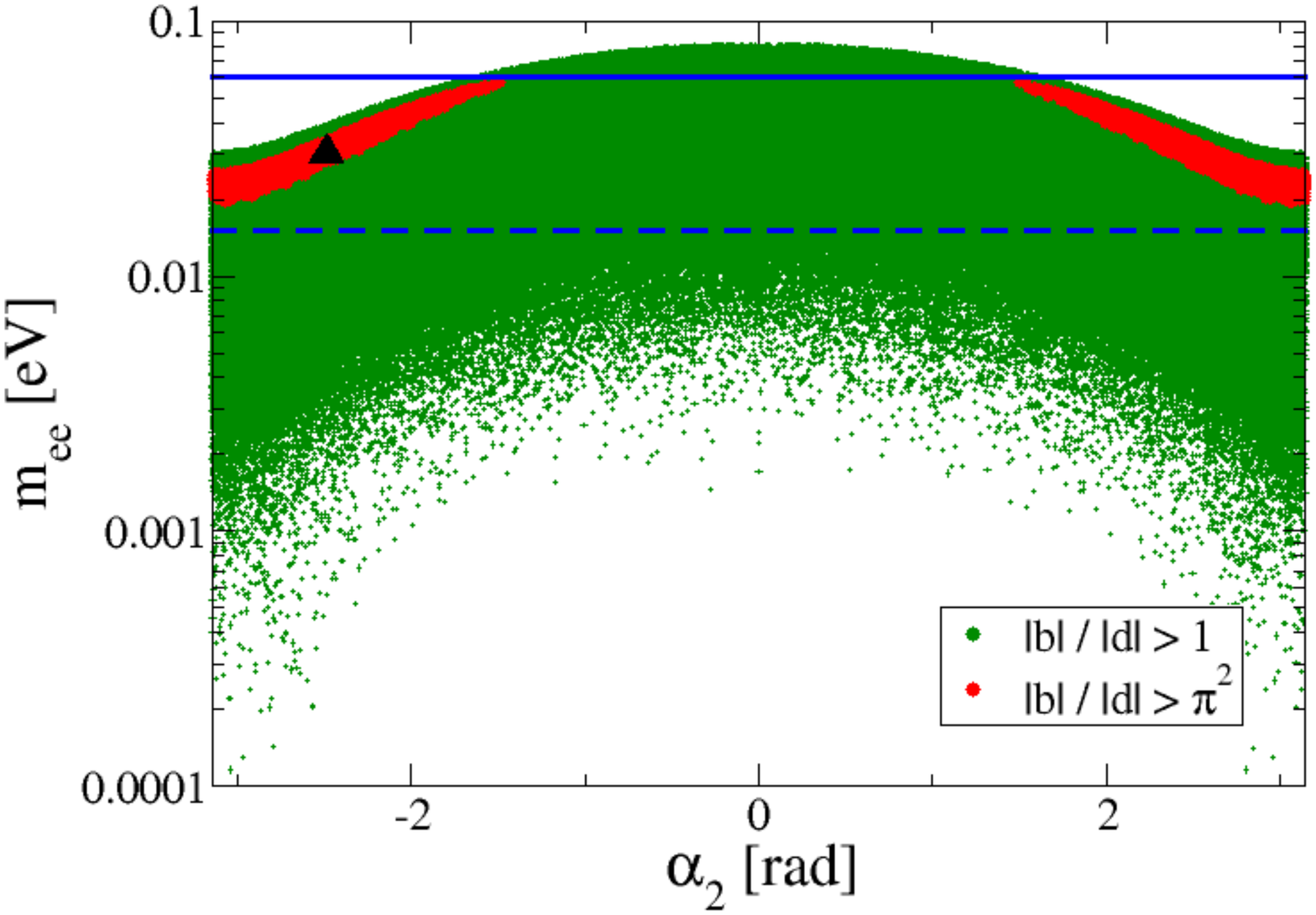}
			\includegraphics[keepaspectratio,scale=0.28]{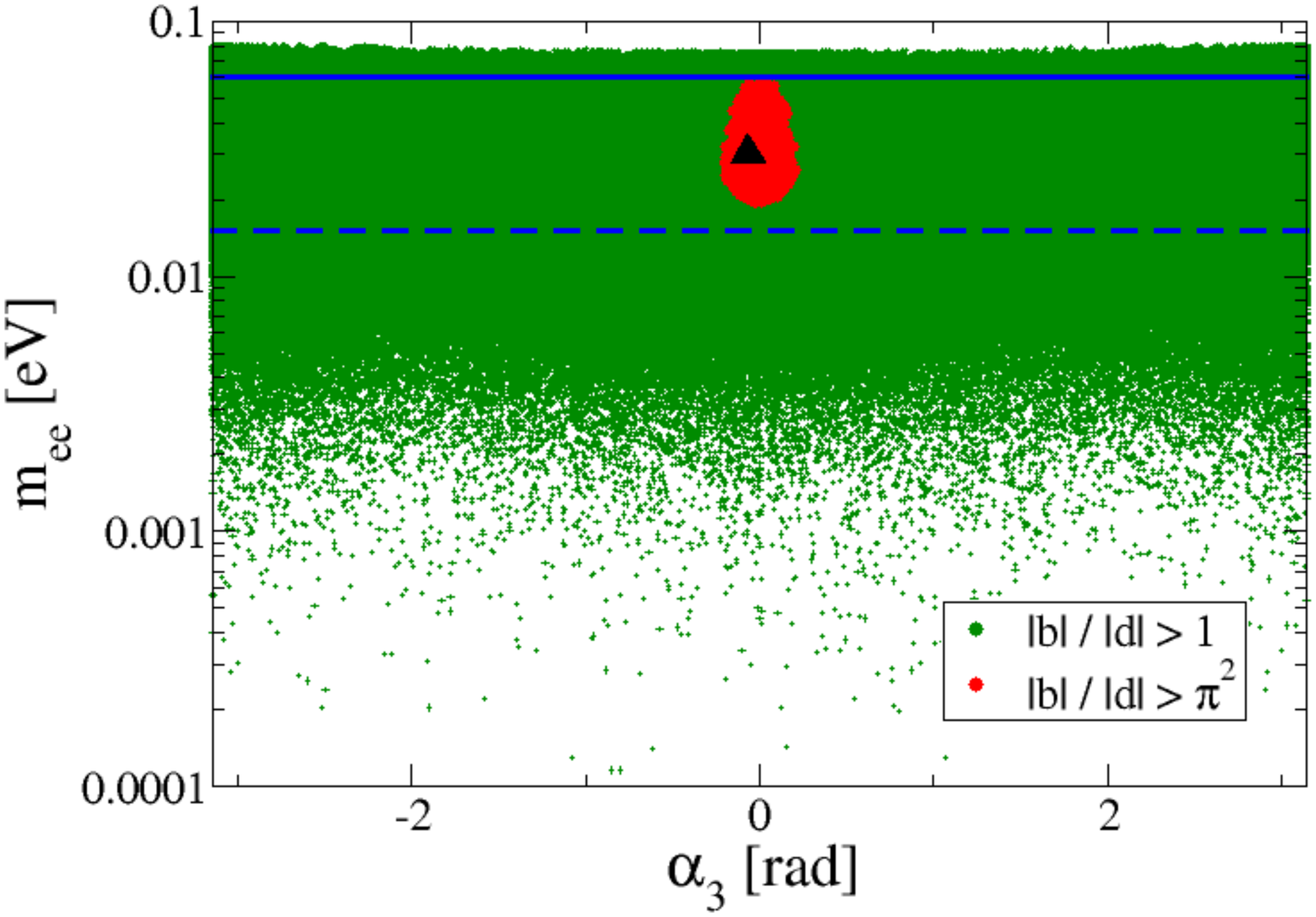}
\caption{$m_{ee}$ versus $m_1$ (upper panel), $\alpha_2$ (lower left panel), $\alpha_3$ (lower right panel) for the NO.
The green points are the same as those in Fig.~\ref{fig:nofighie}.
The red points show the large hierarchy case, $|b|/|d| > \pi^{2}$: the black triangle point indicates the BP$_{{\rm NO}}$.
}
	\label{fig:nofig}
\end{figure}
\begin{figure}[t]
	\centering
			\includegraphics[keepaspectratio,scale=0.28]{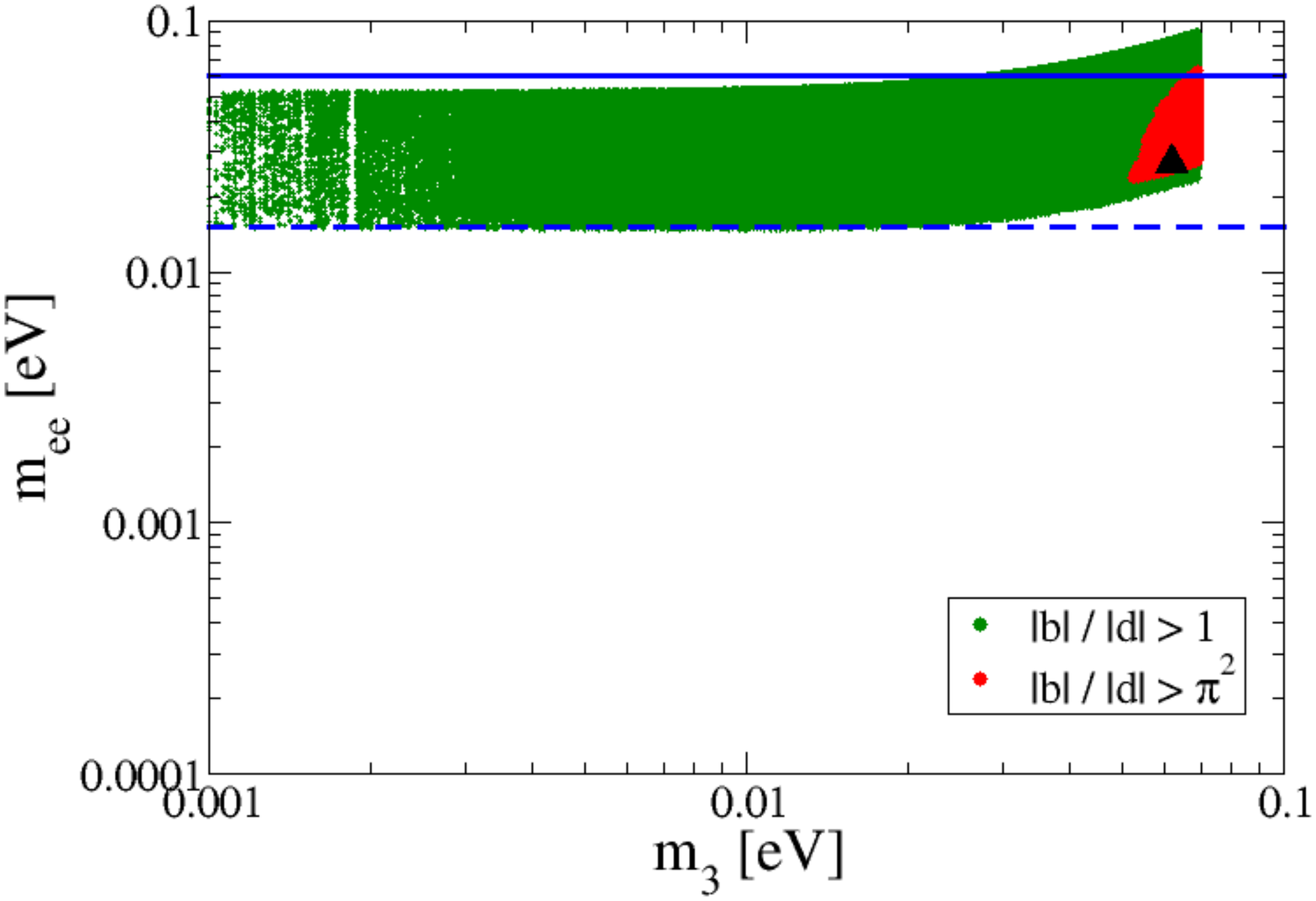}\\
			\includegraphics[keepaspectratio,scale=0.28]{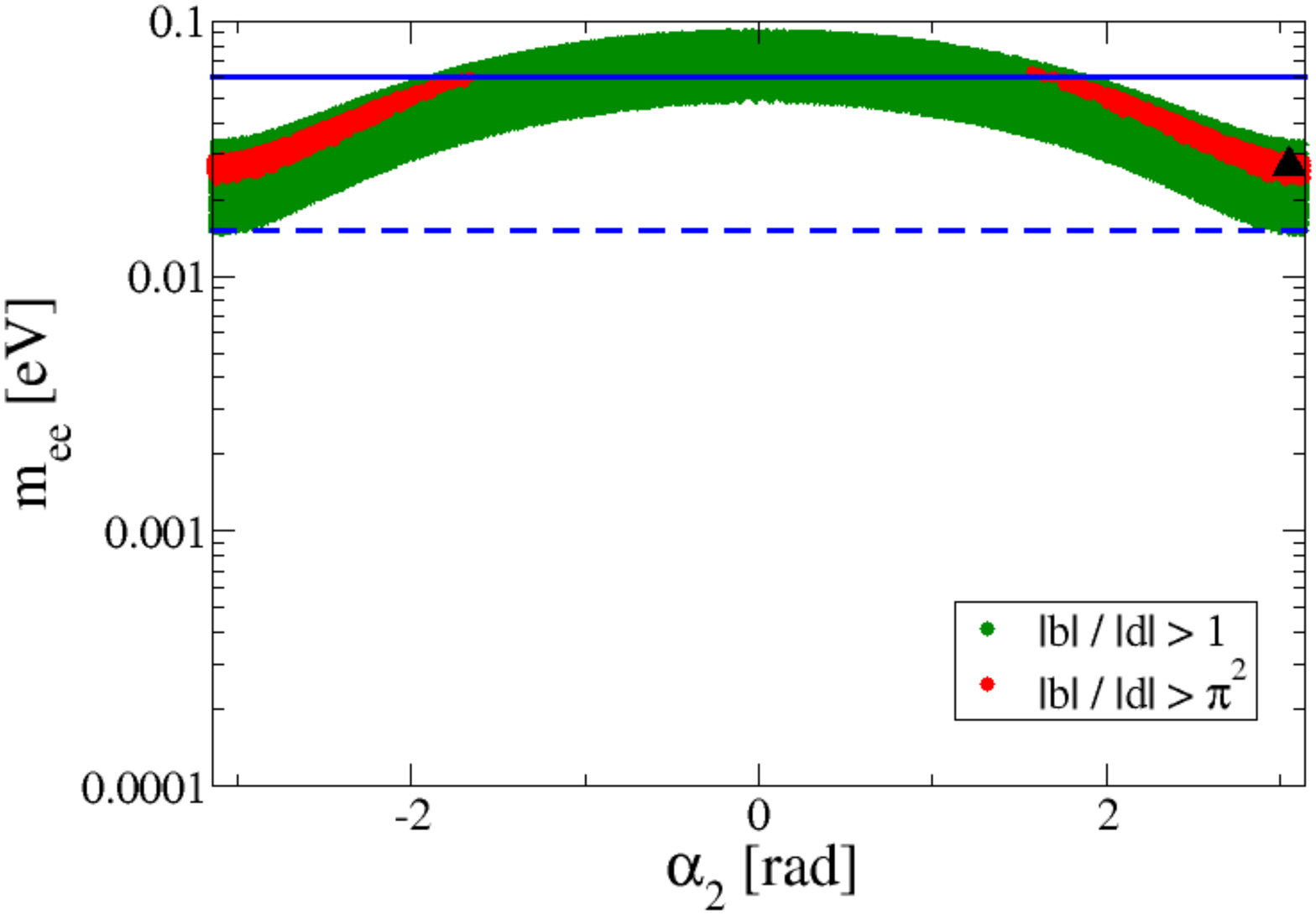}
			\includegraphics[keepaspectratio,scale=0.28]{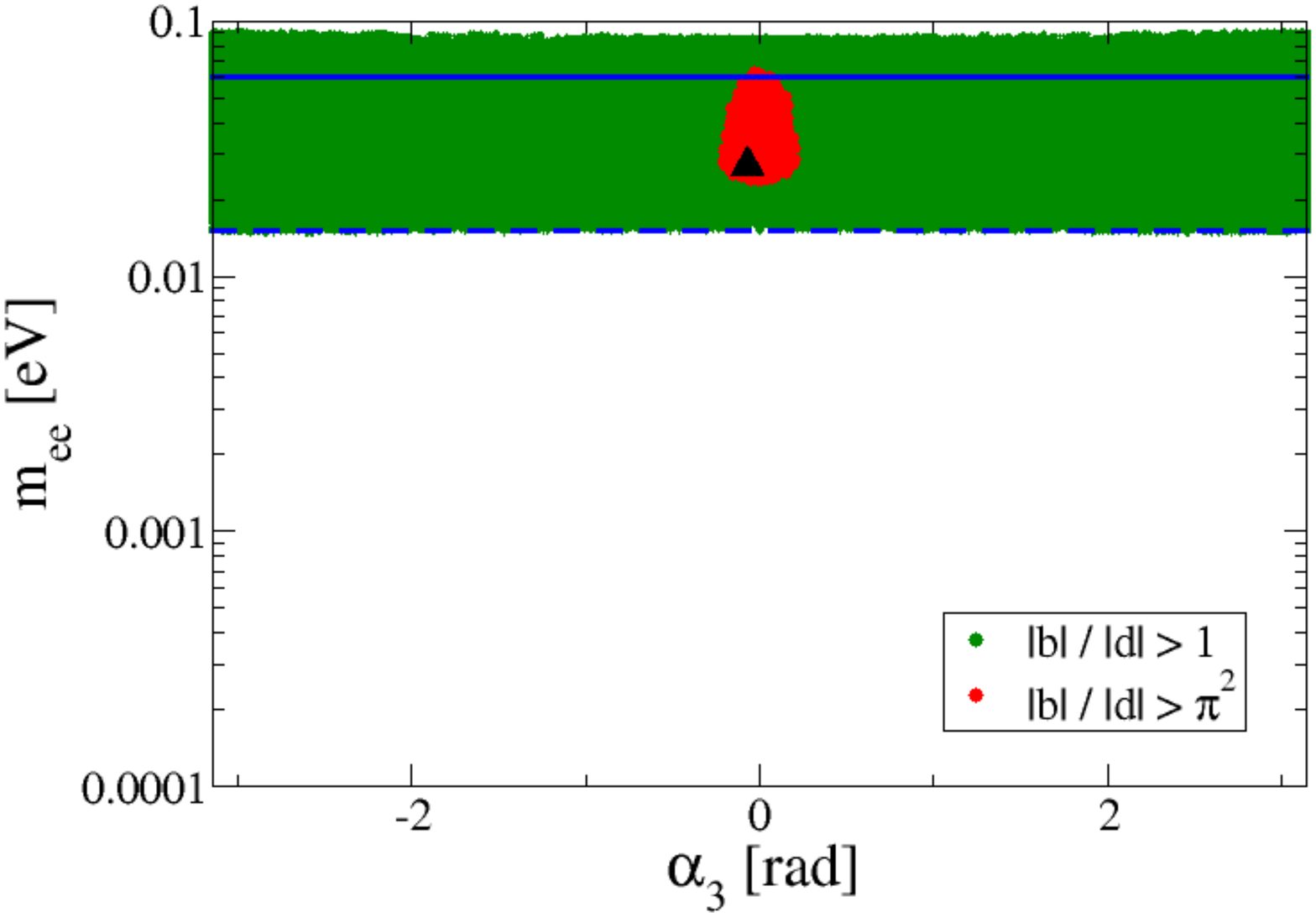}
  \caption{Same as Fig.~\ref{fig:nofig} except for the IO and the lightest neutrino mass $m_{3}$.}
	\label{fig:iofig}
\end{figure}

Next, we discuss the larger hierarchical case with  $|b|/|d| > \pi^{2}$  that can be applied to our model.
We display in Figs.~\ref{fig:nofig} and \ref{fig:iofig} the predictions of $m_{ee}$ for $|b|/|d| > \pi^{2}$ by the red points.
The green points are the predictions for $|b|/|d| > 1$ that
are the same as those in Figs.~\ref{fig:nofighie} and \ref{fig:iofighie}.
We also show the predictions of the BP$_{{\rm NO}}$ and the BP$_{{\rm IO}}$ in our hybrid seesaw model by the black triangles
in Figs.~\ref{fig:nofig} and \ref{fig:iofig}, respectively. 
In Fig.~\ref{fig:nofig}, we can see that the predicted regions for $|b| / |d| > \pi^{2}$ are strictly constrained : 
$ m_{ee} \gtrsim 0.02 ~{\rm eV},~m_{1} \gtrsim 0.06~{\rm eV}, ~ \pi/2 \lesssim |\alpha_{2}|/{\rm rad } \leq \pi,~ |\alpha_{3}|/{\rm rad } \lesssim 0.2$.
Such large $m_{ee}$ is within the sensitivity reach of the next-generation $0\nu\beta\beta$ experiments.
Note that, for the larger hierarchy between the coefficients $|b|$ and $|d|$,  a larger hierarchy between $|a|$ and $|d|$ is expected and thus the Majorana phase $|\alpha_{3}|$ gets closer to zero. In our hybrid seesaw model, the predicted values of the BP$_{{\rm NO}}$ are:
\begin{align}
&m_{ee} \approx 0.030 ~{\rm eV},~m_{1} \approx 0.067~{\rm eV},~\alpha_{2} \approx -2.5 ~{\rm rad},~\alpha_{3} \approx -0.07 ~{\rm rad}
\end{align}
Figure~\ref{fig:iofig} shows the results for the IO, where we can see similar predictions for the red regions with the NO.
 The predicted values of the BP$_{{\rm IO}}$ are: 
\begin{align}
&m_{ee} \approx 0.027 ~{\rm eV},~m_{3} \approx 0.061~{\rm eV },~\alpha_{2} \approx 3.04 ~{\rm rad},~\alpha_{3} \approx -0.07 ~{\rm rad},
\end{align}
where the non-significant CP violations by the Majorana phases are expected.

%
%
\section{Dark matter}
\begin{figure}[t]
\includegraphics[keepaspectratio,scale=0.35]{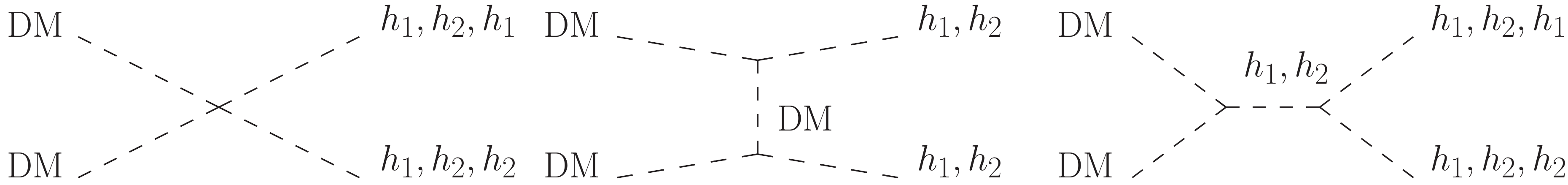}
\caption{Feynman diagrams giving main contributions to the relic abundance.}\label{fig:annihilation}
\end{figure}

The $A_{4}$ flavor symmetry in our model is spontaneously broken to the $Z_{2}$ symmetry via the VEV of the $A_4$ triplet scalar field,
which predicts the DM candidates and we assume that 
the $Z_{2}$-odd scalar field $\eta_{2}^{0}$ is the DM.\footnote{
Such DM (so-called ``discrete DM'') is discussed in Refs. \cite{Hirsch:2010ru,Boucenna:2011tj,Peinado:2011aa,Hamada:2014xha,Mukherjee:2015axj,Bhattacharya:2016rqj,Lamprea:2016egz,Gautam:2019pce}. }
The main annihilation processes of the DM in our scenario are shown in Fig.~\ref{fig:annihilation}.
Note that the processes are almost the same as those in the inert doublet model \cite{LopezHonorez:2006gr}.\footnote{
The scalar fields $\eta_{2}^{0},~\eta_{3}^{0},~A_{2}$ and $A_{3}$ can be probed at the collider experiments via the processes 
that are similar to the inert doublet model \cite{LopezHonorez:2006gr,Barbieri:2006dq,Goudelis:2013uca}.} In our model, the mass splitting between $\eta_2^0$ and $\eta_3^0$ is small because of the small $\lambda_{x_{3}}$ coupling,
so that we also consider the annihilation of $\eta_{3}^{0}$ and the relic density is computed for the sum of $\eta_{2}^{0}$ and $\eta_{3}^{0}$.\footnote{
We note that $\eta_{3}^{0}$ decays into $\eta_{2}^{0}$ through, {\it e.g.}, $\eta_{3}^{0} \to \eta_{2}^{0}\gamma$ after its decoupling.}
The rate of DM annihilation depends on the scalar couplings 
$\lambda_{1},~\lambda_{x_{5}},~\lambda_{x_{6}}, $ and $\lambda_{x_{7}}$, except for the gauge couplings,
where
  $\lambda_{x_{5}} \equiv \lambda_{9} + \lambda_{10} + 2\lambda_{11},~
  \lambda_{x_{6}} \equiv \lambda_{2} + \lambda_{3} + 2\lambda_{4} + \lambda_{5}$, and $\lambda_{x_{7}} \equiv 2\lambda_{2} - \lambda_{3} - 2\lambda_{4} + \lambda_{5} + 2\lambda_{6} + \lambda_{7} + \lambda_{8}$.
For $\sin(\beta - \alpha) = 1$,
the relevant scalar couplings $\lambda_{x_{5}}, \lambda_{x_{6}} $, and $\lambda_{1}$ are given by the masses of the $Z_2$-even neutral scalar fields
 $m_{h_{1}}$, $m_{h_{2}}$ and the mixing angle $\beta$ as
\begin{align}
	\lambda_{x_{5}} &= \frac{m_{h_1}^{2} - m_{h_{2}}^{2}}{v^{2}[ \sin^{2}\beta + \cos^{2}\beta( 1 + 2\sin2\beta)]},\\
	\lambda_{x_{6}} &= \frac{m_{h_{2}}^{2}}{v^{2}\sin^{2}\beta} + \lambda_{x_{5}},\\
	\lambda_{1} &= \frac{\lambda_{x_{5}}\cos2\beta + \lambda_{x_{6}}\sin^{2}\beta}{\cos^{2}\beta}. 
\end{align}
The scalar coupling $\lambda_{x_{7}}$ can be determined to satisfy the relic abundance $\Omega h^{2} \approx 0.12$ \cite{Aghanim:2018eyx}.

The spin-independent cross section of the nucleon is given by
 \begin{align}
	 \sigma_{{\rm SI}} = \frac{1}{\pi}\left( \frac{\lambda_{{\rm DD}}\hat{f}m_{N}}{m_{\eta_2^0}m_{h_1}^{2}}  -  \frac{\lambda'_{{\rm DD}}\hat{f}m_{N}}{m_{\eta_2^0}m_{h_2}^{2}\tan\beta} \right)^{2}\left( \frac{m_{N}m_{{\eta_2^0}}}{m_{N} + m_{{\eta_2^0}}} \right)^{2},
\label{dd}
 \end{align}
 where $\hat{f} \approx 0.3$ is the usual nucleonic matrix element \cite{Ellis:2000ds}, $m_{N}$ is the nucleon mass, $\lambda_{{\rm DD}} = \lambda_{x_{7}}\sin^{2}\beta + \lambda_{x_{5}}\cos^{2}\beta$, and $\lambda'_{{\rm DD}} = \lambda_{x_{7}}\sin2\beta - \lambda_{x_{5}}\sin2\beta$.
 Since the contributions from the $h_1$ and $h_2$ mediations give a relative negative sign, there is a possibility of destructive interference 
for $m_{h_1}\sim m_{h_2}$. 
\begin{figure}[t]
\includegraphics[keepaspectratio,scale=0.35]{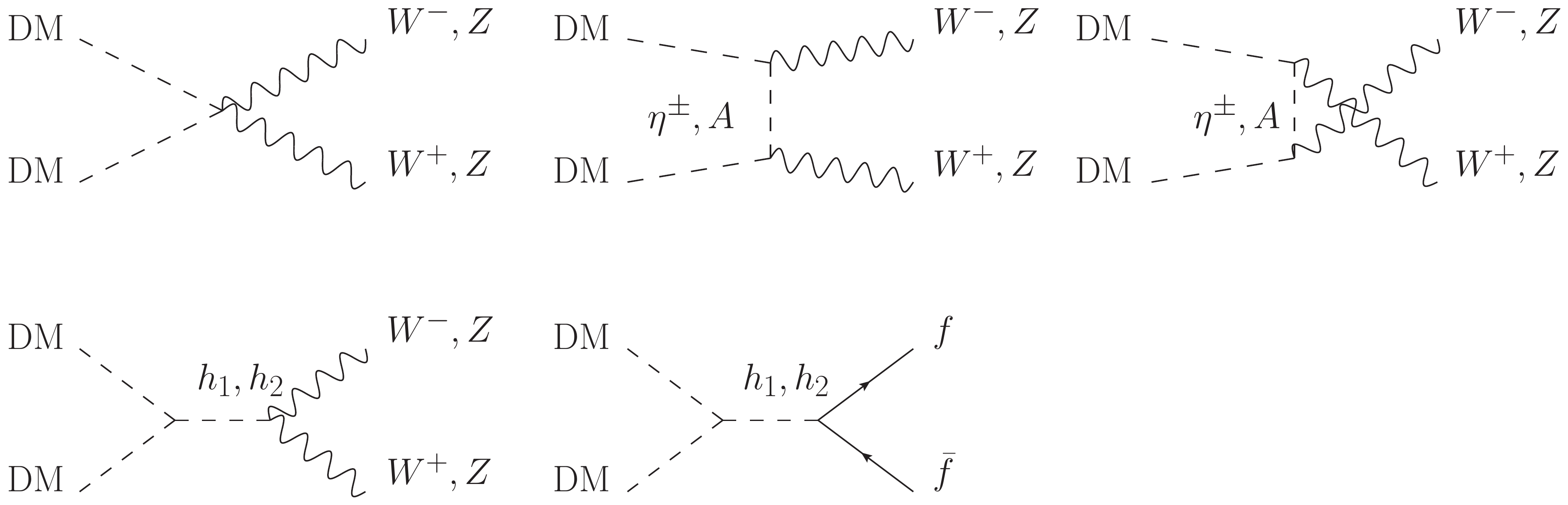}
\end{figure}

\begin{figure}[t]
\vspace{10mm}
	\centering
	\includegraphics[keepaspectratio,scale = 0.3]{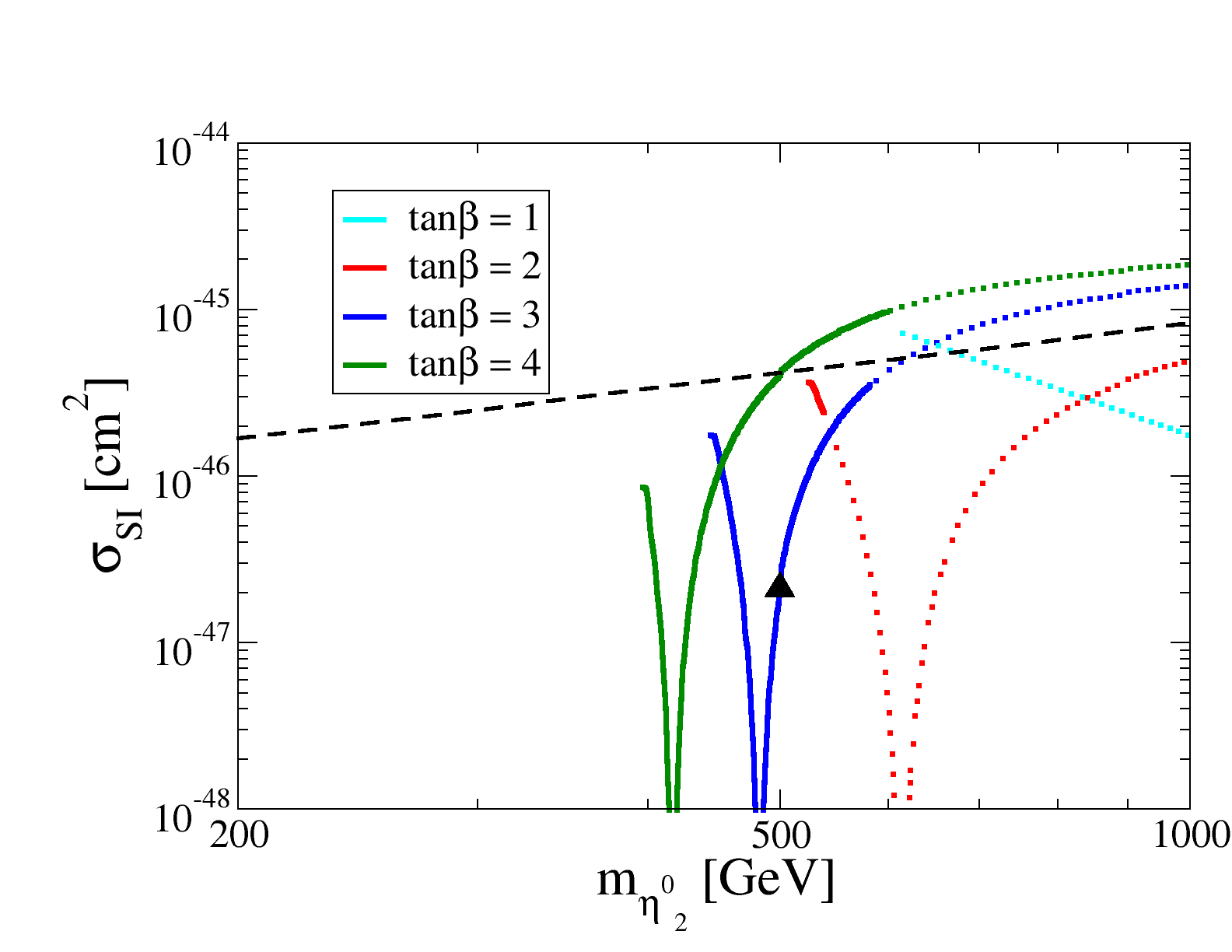}
	\caption{DM mass versus spin-independent cross section. The cyan, red, blue and green lines show the results for $\tan\beta = 1,~ 2$, $3$ and $4$, respectively, for $~m_{h_{2}} = 200 ~{\rm GeV},~m_{H^{\pm}} = m_{A_{1}} = 250~{\rm GeV}$, and $m_{A}=m_{\eta^{\pm}} = m_{\eta^{0}} + 20~{\rm GeV}$. The dotted lines are excluded by the unitarity and the bounded-from-below conditions on the scalar potential \cite{Boucenna:2011tj}.
The region above the black dashed line is excluded by XENON1T \cite{Aprile:2017iyp}.}
	\label{fig:dd}
\end{figure}

In Fig.~\ref{fig:dd}, the spin-independent cross section of DM is shown as a function of the DM mass, where the relic abundance 
of the DM satisfies $\Omega h^{2} \approx 0.12$ \cite{Aghanim:2018eyx}.
Here we have fixed the masses of the $Z_{2}$-even scalar fields
as $m_{h_{2}} = 200 ~{\rm GeV}$ and $~m_{H^{\pm}} = m_{A_{1}} = 250~{\rm GeV}$. 
For the $Z_{2}$-odd  scalar fields, the masses are taken as $m_{A}=m_{\eta^{\pm}} = m_{\eta^{0}} + 20~{\rm GeV}$. 
The cyan, red,  blue and green lines show the results for $\tan\beta = 1,~ 2,~3$, and $4$, respectively,
where the dotted lines are excluded by 
the unitarity condition $\lambda_{i} < 4\pi ~(i = 1 - 14)$ or the bounded-from-below condition on the scalar potential
\cite{Boucenna:2011tj}.
As a reference, we show the prediction of the BP$_{{\rm NO}}$ and the BP$_{{\rm IO}}$ by a black triangle.
The region above the black dashed line is excluded by XENON1T \cite{Aprile:2017iyp}.
In Fig.~\ref{fig:dd}, we can see the cancellations between the contributions of $h_{1}$ and $h_{2}$.
When the DM mass is smaller than about $400~{\rm GeV}$, the relic abundance 
of the DM is smaller than the observed value $\Omega h^{2} < 0.12$. For $\tan\beta \gtrsim 5$, the unitarity condition cannot be satisfied. 
We find that the allowed ranges for the DM mass are 
$m_{\eta_{2}^{0}} \simeq 520-540$ GeV, $~490-580$ GeV, and $400-500$ GeV for $\tan\beta = 2, ~3$, and 4, respectively.
The future sensitivity of the direct detection experiment XENONnT is $\sigma_{{\rm SI}} \sim \mathcal{O}(10^{-47})~{\rm cm^{2}}$ \cite{Aprile:2017iyp},
which can probe our DM scenario.

\section{Conclusion}
We have studied the neutrino mass matrix that is composed of the four flavor structures in Eq.~(\ref{mnu}) based on the $A_{4}$ flavor symmetry,
focusing on the hierarchical flavor structures.
As a model with a large hierarchical structure,
we have proposed a hybrid seesaw model based on the $A_{4}$ flavor symmetry.
In the model, the neutrino masses are generated by the tree-level and the one-loop seesaw mechanisms.
These mechanisms induce different flavor structures and the large hierarchy with $|b|/|d| > \pi^{2}$ via 
the $A_{4}$ triplet fields of the right-handed neutrinos
at the intermediate scale and of the scalar doublet at the electroweak scale.  
The non-zero $\theta_{13}$ is generated by the one-loop seesaw mechanism.
The model predicts a large effective neutrino mass 
$m_{ee} \sim 0.03$ eV, 
which can be tested by future $0\nu\beta\beta$ experiments,
with a Majorana phase $\alpha_3 \sim 0$. 
Furthermore, the $A_4$ flavor symmetry is broken down to the $Z_2$ symmetry in our model and
the $Z_{2}$-odd scalar field $\eta_{2}^{0}$ becomes the DM.
The constraints arising from the DM relic density set its mass in the range of 400 GeV $\lesssim m_{\eta_{2}^{0}} \lesssim 600$ GeV. 
Future direct detection experiments such as XENONnT will be able to access our DM scenario.

We have also performed a model-independent analysis of the neutrino mass matrix in Eq.~(\ref{mnu}), 
particularly for the cases with some hierarchical flavor structures. 
It  has been found that
the hierarchical cases with $|b|/|a| > 1,~|b|/|c| > 1$, and $|d|/|a| > 1$ 
reduce the allowed parameter space and show  characteristic predictions for the Majorana phases.
On the other hand, the hierarchical case with $|b|/|d| > 1$ does not show specific predictions.
However, a larger hierarchy with $|b|/|d| > \pi^{2}$, which can be realized in our hybrid seesaw model, can reduce the predicted parameter region, which will be testable by future $0\nu\beta\beta$ experiments.

\begin{acknowledgments}
 The work of M.~A. is supported in part by the Japan Society for the
Promotion of Sciences Grant-in-Aid for Scientific Research (Grant
No. 17K05412 and No. 20H00160).
\end{acknowledgments}

\end{document}